\documentclass[apsrev4-1,prl,superscriptaddress,longbibliography,twocolumn]{revtex4-1}
\usepackage{color}
\usepackage{amsfonts,amsmath,amssymb}
\usepackage[dvipsnames]{xcolor}
\usepackage{tabularx,hhline,tikz,multirow,array}
\usepackage{bm,enumitem}
\usepackage{dcolumn,stmaryrd}

\usepackage[bookmarksnumbered,pdfpagelabels=true,plainpages=false,colorlinks=true,linkcolor=blue,citecolor=blue,urlcolor=blue]{hyperref}
\usepackage{comment}
\def\beq{\begin{equation}}
\def\eeq{\end{equation}}
\def\bea{\begin{eqnarray}}
\def\eea{\end{eqnarray}}

\DeclareSymbolFont{bbold}{U}{bbold}{m}{n}
\DeclareSymbolFontAlphabet{\mathbbold}{bbold}

\newcommand{\bs}[1]{\boldsymbol{#1}}

\begin{document}
\title{Robustness of topological magnons in disordered arrays of skyrmions}
\author{H. Diego Rosales}
\affiliation{Instituto de F\'isica de L\'iquidos y Sistemas Biol\'ogicos (IFLYSIB), UNLP-CONICET, La Plata, Argentina and Departamento de F\'isica, Facultad de Ciencias Exactas,
Universidad Nacional de La Plata, c.c. 16, suc. 4, 1900 La Plata, Argentina.}
\affiliation{Departamento de Ciencias B\'asicas, Facultad de Ingenier\'ia, UNLP, 1900 La Plata, Argentina}
\email{rosales@fisica.unlp.edu.ar}
\author{Roberto E. Troncoso}
\affiliation{School of Engineering and Sciences, Universidad Adolfo Ib\'a\~nez, Santiago, Chile}
\email{roberto.troncoso.c@uai.cl}
%
%

\begin{abstract}
The effects of disorder on the robustness of topological magnon states of two-dimensional ferromagnetic skyrmions is investigated. It is diagnosed by evaluating a real space topological invariant, the bosonic Bott index (BI). The disorder simultaneously breaks the axially symmetric shape and the crystalline ordering of the skyrmions array. The corresponding magnonic fluctuations and band spectrum are determined in terms of magnetic field and strength of disorder. We observe the closing of the existing band gaps as the individual skyrmions start to occupy random positions. The analysis reveals that topological states (TSs) persist beyond the perturbative limit when skyrmions reach the glassy phase. In addition, the localization of topologically protected edge states is weakened by the disordered skyrmion structure with increasing localization length. Our findings shed light on the physical understanding of the coexistence of disordered magnetic textures and their topological spin fluctuations.
\end{abstract}

\maketitle

{\it Introduction.--} Magnetic skyrmions, swirling topologically stable defects in the magnetization texture, have attracted enormous attention in condensed matter physics during the last decades \cite{Nagaosa2013}. They have been the ground for fundamental research, such as the topological Hall effect \cite{Schulz2012,Nagaosa2013,tome2021topological}, and applications heading towards information technologies such as logic gates and racetrack memory \cite{Sampaio2013}, or neuromorphic computing \cite{Song2020,Yokouchi2022}. The nucleation of skyrmions occurs under specific conditions forming skyrmion crystals (SkX) \cite{Muhlbauer2009,Yu2010,AdamsPRL2011,rosales2015three,gao2020fractional,mohylna2022spontaneous}, and as isolated particle-like entities \cite{rosales2023skyrmion,gomez2024chiral}. In the presence of defects or magnetic disorder, the characteristic crystalline symmetry of SkX is broken, giving rise to quenched structures that lack long-range order, such as glassy phases \cite{HoshinoPRB2018,Chudnovsky2018,Koshibae2018} or disordered structures \cite{SilvaPRB2014,DiazPRB2017,Karube2018,Reichhardt-RMP2022}. On the verge of phase transitions, skyrmions might distribute randomly as crystallization takes place \cite{Huang2020,Esposito2020,Mohanta2020,KarubePRB2020,Meisenheimer2023}. Quantum spin fluctuations, magnons, about the magnetic order of SkX are a niche that harbors exotic phenomena, such as topological helical edge modes \cite{roldanNJP2016,diazPRL2019,diazPRB2020}, topological thermal Hall effect \cite{albarracin2021chiral,AkazawaPRR2022}, pseudo-Landau levels \cite{Weber2022}, and higher-order TSs \cite{HirosawaPRL2020}. Hence, skyrmions set a promising platform to develop topologically protected data based on magnonic fluctuations. 
However, for their successful integration into spintronic devices, understanding the role of structural and magnetic disorders is of paramount importance.

Topologically protected magnonic states have been intensively studied in a wide variety of spin and lattice systems \cite{Zhuo2023}, such as in collinear magnetic phases \cite{TM3,TM7,TM13,HidalgoPRB2020}, or textured magnets \cite{diazPRL2019,roldanNJP2016,diazPRB2020,AkazawaPRR2022}. Nontrivial topology is characterized by topological indices that remain unaltered under smooth deformations and, set by the bulk-boundary correspondence, lead to the robustness of helical edge-states. Thus, the immunity of TSs to disorder, which is ubiquitous and unavoidable in any physical system, is deeply rooted in topological matter \cite{TI1}. In other solid-state systems, such as electrons or photons, the fate of TSs was examined in the presence of strong disorder based on transport properties, giving rise to the topological Anderson insulator phase \cite{TAI1,TAI2}.
 \begin{figure}[thb]
\includegraphics[width=\columnwidth]{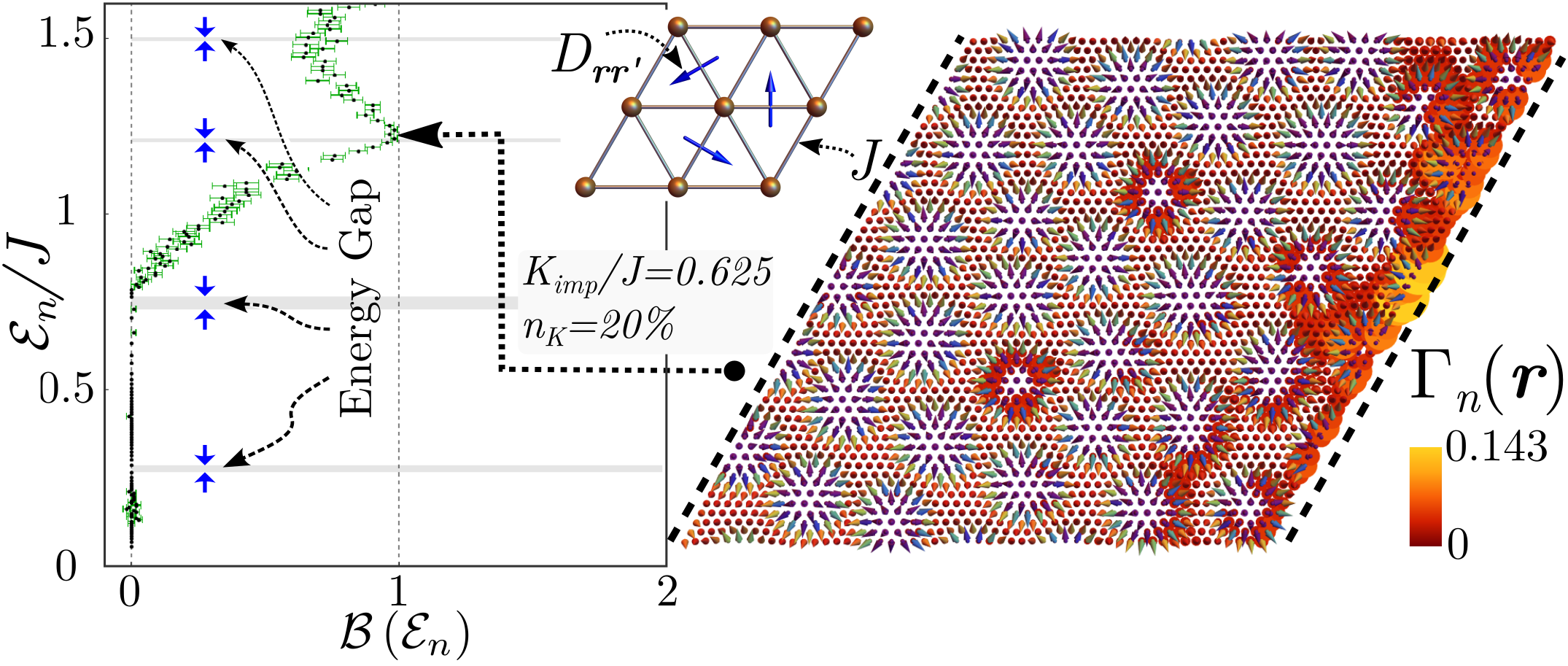}
\caption{Disordered skyrmion arrangement (right panel) and the corresponding Bott index ${\cal B}\left({\cal E}_n\right)$ (left panel) for a system with $K_{\text{imp}}/J = 0.625$ and a 20\% concentration of defects. The spatial distribution of the localized topological magnonic edge-state, $\Gamma_{n}(\bs r)$, is displayed on top of the skyrmions array (on the right panel, dashed lines indicate the open edges of the sample)
.}
\label{fig:fig1}
\end{figure}
Similarly, properties of topological insulators were predicted to occur in amorphous systems \cite{AgarwalaPRL2017,Costa2019,Marsal2020}, with phase transitions monitored by real space topological invariants such as the Chern marker \cite{BiancoPRB2011}, spectral localizer \cite{CerjanPRB2022}, and the BI \cite{Bott3,Toniolo2022}. In bosonic systems, topological magnon phases in collinear ferromagnets show persistence in the limit of strong disorder \cite{WangPRL2020}, where disorder plays an active role in stabilizing the actual phase. Despite these efforts, several queries such as the stability of topological phases, the localization of edge-states, and the transition towards trivial magnonic phases in disordered magnetic textures lacking crystalline order are still underexplored.

In this Letter, we show that topological magnon states prevail in a strongly disordered array of two-dimensional skyrmions. The robustness of topological states is revealed by the evaluation of the bosonic Bott index and the closing of respective band gaps.  The most remarkable fact depicted in Fig.~\ref{fig:fig1} is that the BI remains robust and reaches an integer value, even in this ``dirty'' case with high disorder concentration. The midgap states within the topological (third) gap reveal an ``edge'' state localized on the surface of the lattice. The BI of the first two gaps is zero, indicating trivial bands and the absence of disorder-induced topological transition. The evolution of the topological phase, concerning disorder, is contrasted with skyrmion position correlations, measuring the breakdown of crystalline ordering. The weakening of the localization of topologically protected edge states quantifies a direct effect of the disordered skyrmion structure.

{\it Model.-} We consider a two-dimensional ferromagnetic system with localized spins on a triangular lattice formed by heavy metal atoms having strong spin-orbit coupling. The spin-lattice Hamiltonian is given by
\begin{align}
{\cal H}[\bs S]\nonumber=-\sum_{\langle \bs r,\bs r'\rangle}&\left[J{\bs S}_{\bs r}\cdot{\bs S}_{\bs r'}-{\bs D}_{\bs r\bs r'}\cdot\left({\bs S}_{\bs r}\times{\bs S}_{\bs r'}\right)\right]\nonumber\\
&\qquad-K_{\text{imp}}\sum_{\bs r\in \Omega}\left({S}^{z}_{\bs r}\right)^2-B\sum_{\bs r}{S}^{z}_{\bs r},
\label{eq:Hamiltonian}
\end{align}
where ${\bs S}_{\bs r}$ is a spin operator at site ${\bs r}$ of the lattice with lattice constant $a$. The model takes into account nearest-neighbor ferromagnetic exchange interaction $J>0$, the interfacial Dzyaloshinskii-Moriya (DM) interaction ${\bs D}_{\bs r\bs r'}=D\,\hat{\bs z}\times\hat{\bs d}_{\bs r\bs r'}$ with ${\bs d}_{\bs r\bs r'}={\bs r}-{\bs r}'$, that couples spins located at any pair of sites (see inset in Fig.~\ref{fig:fig1}), and Zeeman interaction due to an external magnetic field ${\bs B}$ along $z$-axis. The disorder is modeled by the easy-axis magnetic anisotropy $K_{\text{imp}}$ randomly distributed in a subset of sites ($\bs r \in \Omega$) at a specific percentage ($n_K$). This simplification is justified as, in real compounds, defects or impurities generally preserve time-reversal symmetry, allowing the representation of disorder as a percentage of sites with non-zero magnetic anisotropy \cite{HoshinoPRB2018}.

\begin{figure}[h!]
\centering
\includegraphics[width=8.5cm]{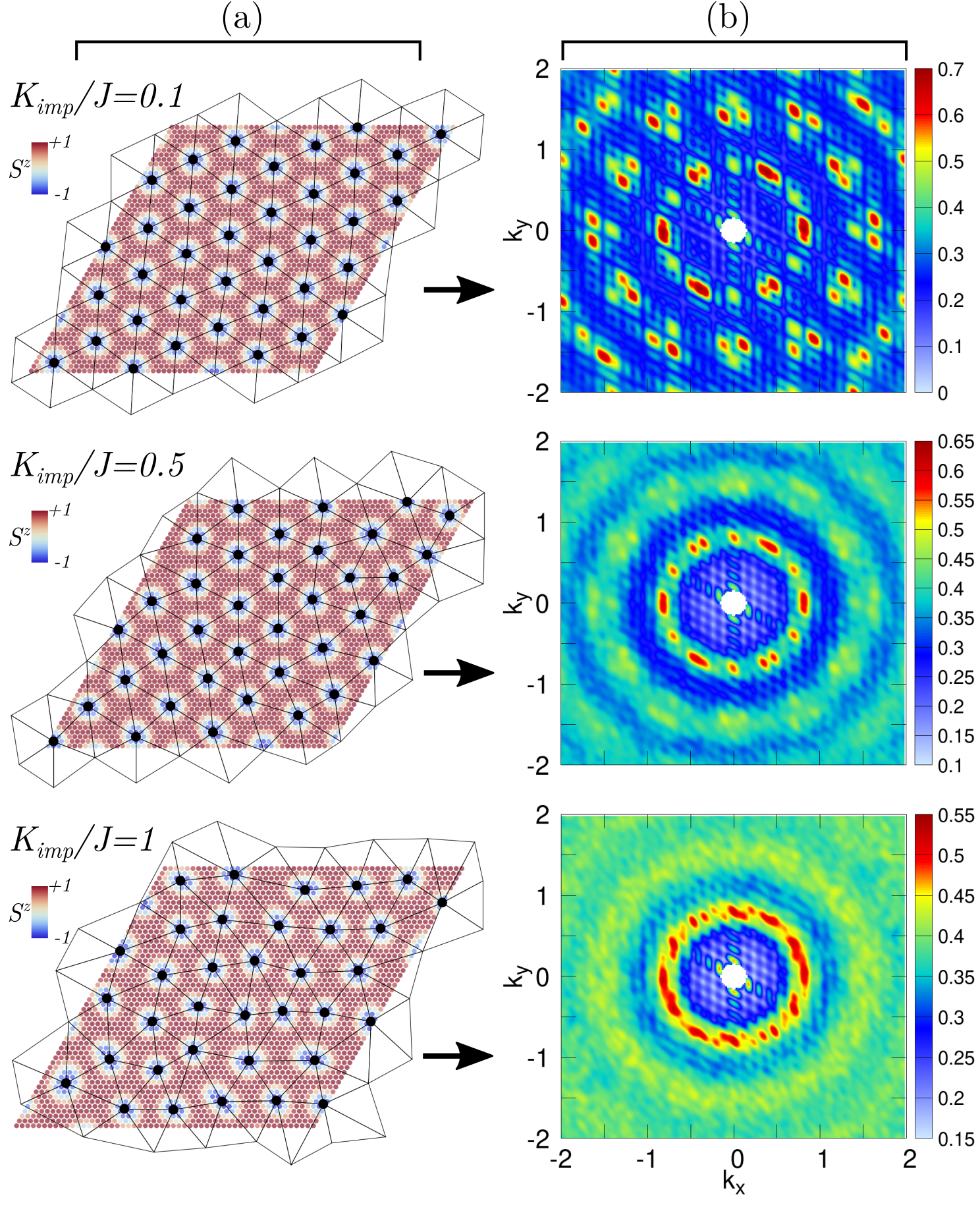}
\caption{(a) Arrays of representative skyrmion configurations for various strengths of disorder $K_{\text{imp}}/J = 0.1$, $0.5$, and $1$ and a 20\% concentration of defects ($n_K$). (b)  The corresponding static structure factor, $S({\bs k})$, for the positions of skyrmions. It was determined averaging over 200 independent disorder realizations.}
\label{fig:fig2}
\end{figure}

The classical spin ground-state of the model in Eq.~(\ref{eq:Hamiltonian}) is obtained by performing extensive Monte Carlo simulations first and then time evolution according to the Landau-Lifshitz-Gilbert equation \cite{liu2006handbook}. All the results are averaged over $200$ independent simulations to estimate the statistical errors (simulation details are given in the SM). 
 \begin{figure*}[thb]
\includegraphics[width=\textwidth]{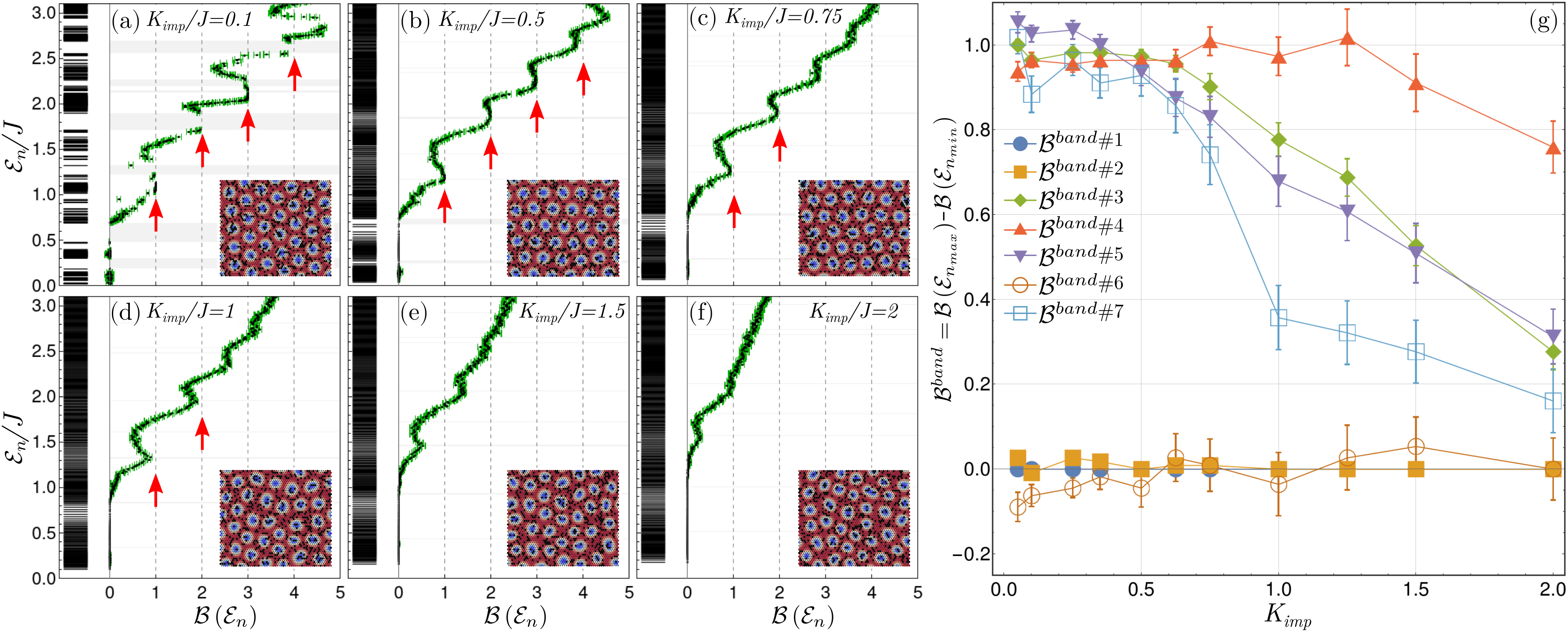}
\caption{Evolution of the Bott index ${\cal B}\left({\cal E}_n\right)$ for various strengths of $K_{\text{imp}}$ (the columns of black lines (resembling a barcode) is depicted alongside, representing the eigen-energies of the problem). Panels (a)-(f): In all cases, at low energies, the first two bands are non-topological, reflected in a zero value of the Bott index. As energy increases, the subsequent bands become topological with an integer index (${\cal B}=1$). Panel (g): Interestingly, the topological character remains robust even in the regime of strong disorder ($K_{\text{imp}}\simeq1$ and $n_K=20\%$ of impurities. Here we plot the Bott index associated with a specific band defined as ${\cal B}^{band}={\cal B}\left({\cal E}_{n_{max}}\right)-{\cal B}\left({\cal E}_{n_{min}}\right)$, where $n_{min} (n_{max})$ correspond to the index of the lower (upper) edge of a given energy band.)}
\label{fig:fig3}
\end{figure*}
%
In the disorder-free model, the application of a magnetic field $B$ leads to the nucleation of SkXs at low temperature ($T \ll J$) \cite{diazPRB2020}, being the skyrmion size fixed by the ratio, $D/J$. To isolate the effects of disorder on the SkX phase, we present simulations assuming $D/J = 1$ and $B/JS=0.6$. For various disorder strengths, $K_{\text{imp}}$, the axially symmetric shape and the crystalline ordering of skyrmions are broken (see Fig.~\ref{fig:fig2}(a)). The pinning of skyrmions in random positions is characterized by the static structure factor $S({\bs k})=N_{sk}^{-1}\left|\sum^{N_{sk}}_{j}e^{i{\bs k}\cdot{\bs r}_j}\right|^2$, with $N_{sk}$ the total number of skyrmions and $\bs r_j$ their positions \footnote{In presence of disorder the shape of each skyrmion is distorted and thus, we assume for their positions the location of the spin pointing downwards in the magnetic structure.}. At Fig.~\ref{fig:fig2}(b) (and Fig.~S1 in the Supplemental Material (SM)) the evaluation of $S({\bs k})$ (averaged over independent realizations) shows that the triangular crystalline order, existing at null or small disorder, fades away for increasing impurity strengths. Moreover, for large enough disorder an emerging strongly peaked ring, together with a widened of the Bragg peaks, indicates a pinning-depinning transition and the rising of a glassy phase of skyrmions \cite{Chudnovsky2018,HoshinoPRB2018}. In this regime, skyrmions lack long-range order and, on average, are separated by a distance that is similar to the disorder-free system.  The radial distribution function, denoted as $g(r)$ where $r$ represents the distance between skyrmions, is illustrated in Fig.~S2 (see SM for details). For $K_{\text{imp}}<0.5$, it exhibits characteristic peaks that align with the quasi-long-range positional ordering typical of a solid-like phase, as reported in previous studies \cite{nishikawa2019solid,Huang2020,zazvorka2020skyrmion,rosales2023skyrmion}. Conversely, for values of $K_{\text{imp}}>0.5$, a well-defined oscillatory pattern emerges, indicative of a transition towards a glassy phase within the skyrmion system \cite{raftrey2023ordering}. This transition results in the skyrmions being kinetically frozen into a glassy state, thereby altering their dynamics.

{\it Magnon spectrum.-} To describe quantum spin fluctuations about the ground state of magnetic skyrmions, we determine the Hamiltonian of non-interacting magnonic excitations using Holstein-Primakoff (HP) bosons \cite{HolsteinPR1940}. Due to the noncollinearity of skyrmion textures, it is convenient to reorient the spin quantization axis along the direction of the classical ground state. We define a local coordinate system at position $\bs r$ by the orthonormal basis $\{\bs{\hat{e}}^1_{\bs r},\bs{\hat{e}}^2_{\bs r},\bs{\hat{e}}^3_{\bs r}\}$, such that $\langle {\bs S_{\bs r}}\rangle=S\,\bs{\hat{e}}^{3}_{\bs r} $. Thus, the rotation of the spin operator is accomplished by ${\bs S}_{\bs r}=\sum_{\alpha}\bs{\hat{e}}^{\alpha}_{\bs r}{\cal S}^{\alpha}_{\bs r}$ and the HP mapping reads ${\cal S}^{+}_{\bs r}=\left(2S-a^{\dagger}_{\bs r}a_{\bs r}\right)^{1/2}a_{\bs r}$, ${\cal S}^{-}_{\bs r}=a^{\dagger}_{\bs r}\left(2S-a^{\dagger}_{\bs r}a_{\bs r}\right)^{1/2}$ and ${\cal S}^{3}_{\bs r}=S-a^{\dagger}_{\bs r}a_{\bs r}$. 
Moreover, since the crystalline order of skyrmions is broken by the presence of disorder, the Brillouin zone of magnonic excitations is ill-defined. Thus, expanding the spin Hamiltonian as a series in $1/S$, the real space magnon Hamiltonian reads ${\cal H}_m={{\cal E}_0}+\frac{1}{2}{\bs a}^{\dagger}\cdot {\bf {\cal M}}\cdot{\bs a}$, where the vector of operators is ${\bs a}=\left(a_{\bs r},a^{\dagger}_{\bs r}\right)^T$ with $\bs r$ running over all $N$ sites, ${\cal E}_0$ is the classical energy that depends on the disordered spin configuration, and ${\bf {\cal M}}$ is a matrix determined by the magnetic texture and microscopic couplings, defined at the SM. The bosonic Hamiltonian is diagonalized by the Bogoliubov transformation $(a_{\bs r},a^{\dagger}_{\bs r})^T=\Psi_{\bs r\bs r'}(\alpha_{\bs r'},\alpha^{\dagger}_{\bs r'})^T$, with $\Psi$ a paraunitary transformation that satisfy $\Psi^{\dagger}\zeta\Psi=\zeta$ in order to preserve the commutation relation $\left[{\bs \alpha},{\bs \alpha}^{\dagger}\right]=\mathbb{I}_{N\times N}\otimes\sigma_z=\zeta$ for bosonic operators \cite{colpa1978}. This transformation leads to a magnon Hamiltonian of the form ${\cal H}_{m}={\cal E}_0+S\sum_{n}{\cal E}_{n}(\alpha^{\dagger}_{n}\alpha_{n}+1/2)$ with ${\cal E}_{n}$ the $n^{\text{th}}$-eigen-energy. It is worth noting that the calculation of the spectra is carried out for each realization of disorder, then the obtained eigen-energies and -states are averaged among all the samples.

{\it Topological phases and Bott index.-} The topology of magnonic bands is characterized by the evaluation of the bosonic BI \cite{WangPRL2020}. It is a real-space topological index that discerns whether or not a system is topological and is shown to be equivalent to the Chern number, in systems with translational invariance and in the thermodynamic limit 
\cite{Bott3,Toniolo2022}. For a set of eigenvalues $\left\{{\cal E}_n\right\}$, it is defined as ${\cal B}\left({\cal E}_n\right)=\text{Im}\left[\text{Tr}\left[\text{log}\left(VUV^{\dagger}U^{\dagger}\right)\right]\right]/2\pi$, where the matrices $V$ and $U$ are defined by
\begin{gather}
Pe^{i\pi X}P=\Psi\zeta \left(\begin{matrix}0 & 0 \\0 & U \end{matrix}\right)\Psi^\dagger\zeta,\\ Pe^{i\pi Y}P=\Psi\zeta \left(\begin{matrix}0 & 0 \\0 & V \end{matrix}\right)\Psi^\dagger\zeta,
\end{gather}
where $P=\Psi\zeta\Gamma_{\cal N}\Psi^\dagger\zeta$ is the projector on states $\left\{{\cal E}_n\right\}$, $X$ and $Y$ are the position operators, and the diagonal matrix $[\Gamma_{\cal N}]_{nn'}=\gamma\delta_{nn'}$, with $\gamma=1$ when $1\leq n\leq{\cal N}$, and $\gamma=0$ for ${\cal N}<n$ \footnote{For fermionic systems the metric $\zeta$ reduces to the identity and the definition in the main text returns to the electronic Bott index}. $\mathcal{B}$ is an integer as long as $VUV^\dagger U^\dagger$ is non-singular, and in particular, $\mathcal{B}=0$ when $U$ and $V$ commute. In a clean system characterized by well-defined energy gaps, the BI for each band separated by these gaps is easily discernible. However, in the disordered skyrmion phase, the BI has to be calculated and averaged across various disorder realizations.

The Bott index and magnon spectra are shown in Fig.~\ref{fig:fig3} for different disorder strengths and, averaged over $200$ disorder realizations. As the eigen-energies increase, the BI is displayed for a high percentage of impurities ($n_K=20\%$). At null disorder, the SkX displays the typical triangular arrangement of skyrmions \cite{roldanNJP2016,diazPRB2020}, while the presence of disorder results in the pinning of skyrmions in random positions, see Figs.~\ref{fig:fig2} and \ref{fig:fig3}. At small disorder strength, $K_{\text{imp}}/J=0.1$, the existing topological gaps are identified by the plateaus where the BI reaches integer values, matching previous studies based on the evaluation of Chern number in periodic arrangements \cite{diazPRB2020}. Remarkably, at higher impurity levels (up to $n_K=20\%$) and larger $K_{\text{imp}}$ values, the topological nature of the disordered skyrmion system persists, see Fig.~\ref{fig:fig3}(a-f). This robustness is manifested in two ways: firstly, through a preserved integer BI (indicated by the dashed arrows), with a narrower distribution of plateaus as the topological gap closes \cite{RostamiPRB2017}. Secondly, by the presence of magnonic states localized on the edge of the lattice, as shown in Fig.~\ref{fig:fig1} (right panel), by an accurate choice of open boundary conditions. For sufficiently large values of $K_{\text{imp}}>1$, the topological character is lost since the BI takes non-integer values. This is observed in Figs.~\ref{fig:fig3}(e) and (f), where the BI remains below integer values for energies located within the gaps. Interestingly, the vanishing of TSs of magnons is correlated with the development of a glassy skyrmion phase, evidenced by the static structure factor (see Fig.~\ref{fig:fig2} and Fig.~S1 in the SM). 

\begin{figure}[thb]
\includegraphics[width=\columnwidth]{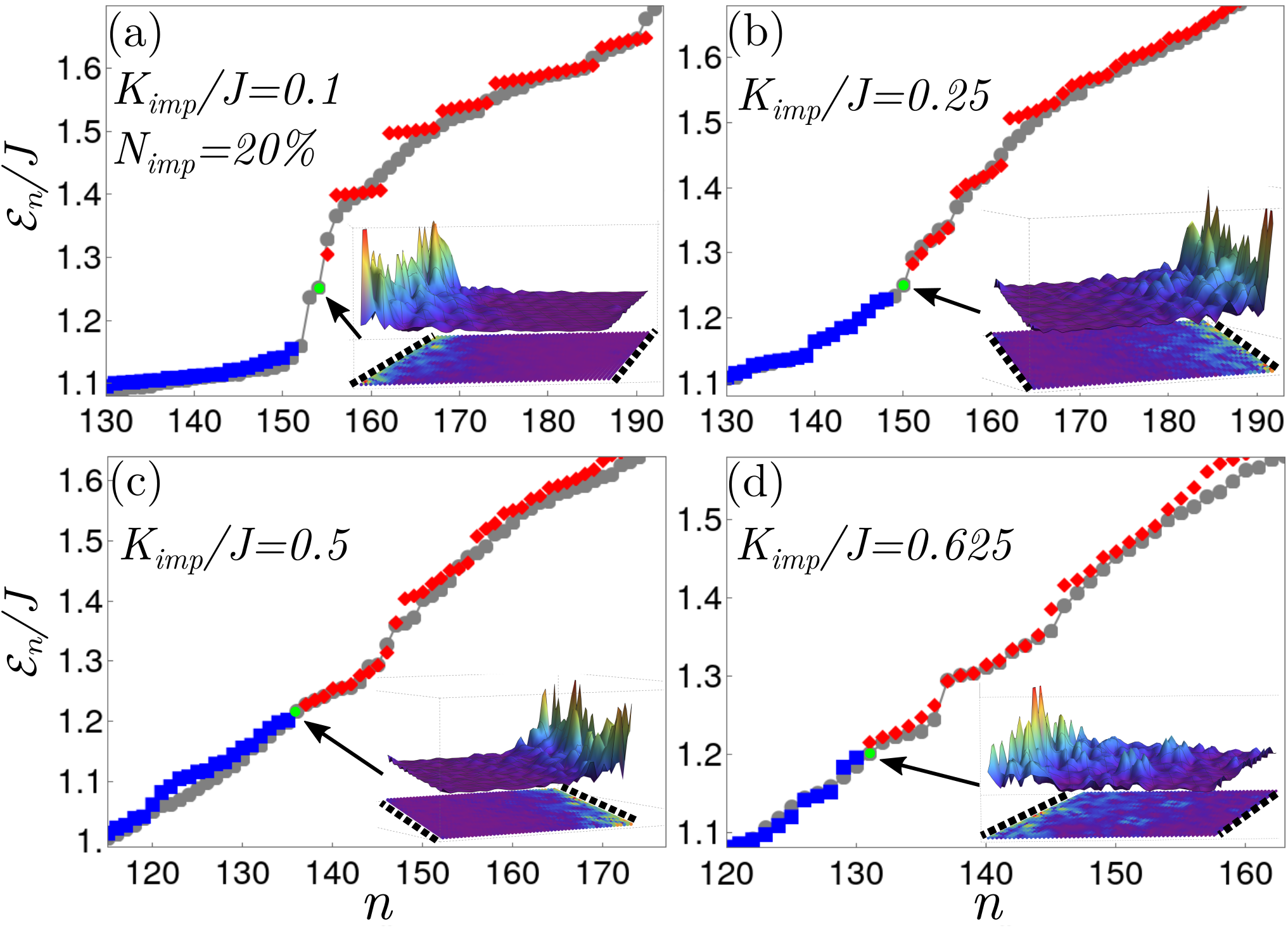}
 \caption{Energy spectrum and real space characterization of edge states for different strengths of disorder. Around the first topological gap the right- and (left-)moving states, encircled in green, are localized near the respective boundary until the gap is closing. The magnonic contribution of the edge states wave function, $\Gamma_{L,R}({\bs r})$, is displayed at the insets.}
\label{fig:fig4}
\end{figure}

{\it Edge-states and localization.-} We now discuss the existence of edge states, established by the bulk-edge correspondence, in the presence of disorder. For a strip geometry, we consider finite size along $x$ axis and periodic boundary conditions along the perpendicular direction.
In Fig.~\ref{fig:fig4}, the energy spectrum around the first topological gap is shown by increasing the disorder. The real space characterization of magnonic states is determined by the overlap, $\Gamma_{n}({\bs r})=\left|\langle GS|a_{\bs r}\alpha^{\dagger}_{n}|GS\rangle\right|^2$, between the local excitation and the eigenstates. The excited single-magnon state is $|n\rangle=\alpha^{\dagger}_{n}|{GS}\rangle$ for the $n^{\text{th}}$ band, with $|{GS}\rangle$ the ground-state of the magnon Hamiltonian. The magnonic contribution of the right- and (left-)moving edge states wave function, $\Gamma_{L,R}({\bs r})$, is displayed in the insets of Fig.~\ref{fig:fig4}. As expected, these states localize near the respective boundary. However, as the impurity strength approaches $K_{\text{imp}}/J=1$ the gap closes and the wave function spreads out through the bulk, characterized by an increasing localization length. The weakening of the magnonic edge states sets the onset of their delocalization and therefore, the topological phase transition to trivial states. This phenomenon is also evident in other topological gaps, as seen in Figs.~S3 and S4 of the SM, where is clear that topological edge states remain localized for strong disorders where a glassy phase of skyrmions is expected. To further complement the analysis, Fig.~S5 shows the BI for various values of $K_{imp}$ and $n_K$, revealing noteworthy robustness of the Bott index even under strong disorder regimes.

{\it Discussions.-} Our work extends the understanding of topological materials by exploring the interplay between disorder and topology of magnonic excitations of skyrmion systems. Skyrmion lattices, characterized by a non-zero topological charge, support magnonic bands with topologically protected edge states. We demonstrate that these topological features persist even when the crystalline order of skyrmions is disrupted by localized magnetic impurities. This robustness is quantified by the BI, which remains quantized even for substantial values of $K_{imp}\sim J$ and an impurity concentration of $n_K=20\%$, indicating the existence of a topological phase. The presence of disorder weakens the topological character in two ways. First, the topological gaps close, and second, the topological edge states progressively delocalize into the bulk, as the disorder strength increases. The delocalization signifies a transition towards a trivial magnonic phase. Interestingly, the loss of topological character coincides with the emergence of a glassy skyrmion phase, as identified by the static structure factor. Experimental access to magnonic edge states might be provided by magnon dispersion measurements using inelastic neutron scattering \cite{chisnell2015topological} or optically imaging \cite{VinasPRL2022}. However, the spatial localization effects induced by disorder and their correlations might be detected via 
the state-of-the-art NV centers techniques \cite{Zhou2021}. In setups like the layered systems Pt/FM/Au/FM/Pt \cite{hrabec2017current} or Ta/Pt/Co/MgO/Ta films \cite{juge2019current}, where impurities are responsible for pinning of skyrmions, might serve as ideal platforms for experimentally analyzing the effects on topologically protected edge states.

In summary, our work paves the way for future research on magnon transport in disordered magnetic textures. The interplay between disorder-induced localization and topological protection offers promising avenues for designing novel systems with enhanced functionalities, e.g., magnon-based spin logic devices or chiral waveguides with superior robustness against defects. These findings can encourage material science experts to address the challenges of realizing these possibilities in real materials.

R.E.T thanks funding from Fondecyt Regular 1230747.
H.D.R. acknowledge financial support from CONICET (PIP 2021-11220200101480CO), Agencia I+D+i (PICT 2020 Serie A 03205) and SECyT-UNLP (I+D X893).

\bibliographystyle{apsrev4-1}
\bibliography{disordered-SkX}

\begin{thebibliography}{63}%
\makeatletter
\providecommand \@ifxundefined [1]{%
 \@ifx{#1\undefined}
}%
\providecommand \@ifnum [1]{%
 \ifnum #1\expandafter \@firstoftwo
 \else \expandafter \@secondoftwo
 \fi
}%
\providecommand \@ifx [1]{%
 \ifx #1\expandafter \@firstoftwo
 \else \expandafter \@secondoftwo
 \fi
}%
\providecommand \natexlab [1]{#1}%
\providecommand \enquote  [1]{``#1''}%
\providecommand \bibnamefont  [1]{#1}%
\providecommand \bibfnamefont [1]{#1}%
\providecommand \citenamefont [1]{#1}%
\providecommand \href@noop [0]{\@secondoftwo}%
\providecommand \href [0]{\begingroup \@sanitize@url \@href}%
\providecommand \@href[1]{\@@startlink{#1}\@@href}%
\providecommand \@@href[1]{\endgroup#1\@@endlink}%
\providecommand \@sanitize@url [0]{\catcode `\\12\catcode `\$12\catcode
  `\&12\catcode `\#12\catcode `\^12\catcode `\_12\catcode `\%12\relax}%
\providecommand \@@startlink[1]{}%
\providecommand \@@endlink[0]{}%
\providecommand \url  [0]{\begingroup\@sanitize@url \@url }%
\providecommand \@url [1]{\endgroup\@href {#1}{\urlprefix }}%
\providecommand \urlprefix  [0]{URL }%
\providecommand \Eprint [0]{\href }%
\providecommand \doibase [0]{http://dx.doi.org/}%
\providecommand \selectlanguage [0]{\@gobble}%
\providecommand \bibinfo  [0]{\@secondoftwo}%
\providecommand \bibfield  [0]{\@secondoftwo}%
\providecommand \translation [1]{[#1]}%
\providecommand \BibitemOpen [0]{}%
\providecommand \bibitemStop [0]{}%
\providecommand \bibitemNoStop [0]{.\EOS\space}%
\providecommand \EOS [0]{\spacefactor3000\relax}%
\providecommand \BibitemShut  [1]{\csname bibitem#1\endcsname}%
\let\auto@bib@innerbib\@empty
\bibitem [{\citenamefont {Nagaosa}\ and\ \citenamefont
  {Tokura}(2013)}]{Nagaosa2013}%
  \BibitemOpen
  \bibfield  {author} {\bibinfo {author} {\bibfnamefont {N.}~\bibnamefont
  {Nagaosa}}\ and\ \bibinfo {author} {\bibfnamefont {Y.}~\bibnamefont
  {Tokura}},\ }\href {http://dx.doi.org/10.1038/nnano.2013.243} {\bibfield
  {journal} {\bibinfo  {journal} {Nature Nanotechnology}\ }\textbf {\bibinfo
  {volume} {8}},\ \bibinfo {pages} {899–911} (\bibinfo {year}
  {2013})}\BibitemShut {NoStop}%
\bibitem [{\citenamefont {Schulz}\ \emph {et~al.}(2012)\citenamefont {Schulz},
  \citenamefont {Ritz}, \citenamefont {Bauer}, \citenamefont {Halder},
  \citenamefont {Wagner}, \citenamefont {Franz}, \citenamefont {Pfleiderer},
  \citenamefont {Everschor}, \citenamefont {Garst},\ and\ \citenamefont
  {Rosch}}]{Schulz2012}%
  \BibitemOpen
  \bibfield  {author} {\bibinfo {author} {\bibfnamefont {T.}~\bibnamefont
  {Schulz}}, \bibinfo {author} {\bibfnamefont {R.}~\bibnamefont {Ritz}},
  \bibinfo {author} {\bibfnamefont {A.}~\bibnamefont {Bauer}}, \bibinfo
  {author} {\bibfnamefont {M.}~\bibnamefont {Halder}}, \bibinfo {author}
  {\bibfnamefont {M.}~\bibnamefont {Wagner}}, \bibinfo {author} {\bibfnamefont
  {C.}~\bibnamefont {Franz}}, \bibinfo {author} {\bibfnamefont
  {C.}~\bibnamefont {Pfleiderer}}, \bibinfo {author} {\bibfnamefont
  {K.}~\bibnamefont {Everschor}}, \bibinfo {author} {\bibfnamefont
  {M.}~\bibnamefont {Garst}}, \ and\ \bibinfo {author} {\bibfnamefont
  {A.}~\bibnamefont {Rosch}},\ }\href {http://dx.doi.org/10.1038/nphys2231}
  {\bibfield  {journal} {\bibinfo  {journal} {Nature Physics}\ }\textbf
  {\bibinfo {volume} {8}},\ \bibinfo {pages} {301–304} (\bibinfo {year}
  {2012})}\BibitemShut {NoStop}%
\bibitem [{\citenamefont {Tom{\'e}}\ and\ \citenamefont
  {Rosales}(2021)}]{tome2021topological}%
  \BibitemOpen
  \bibfield  {author} {\bibinfo {author} {\bibfnamefont {M.}~\bibnamefont
  {Tom{\'e}}}\ and\ \bibinfo {author} {\bibfnamefont {H.~D.}\ \bibnamefont
  {Rosales}},\ }\href {\doibase https://doi.org/10.1103/PhysRevB.103.L020403}
  {\bibfield  {journal} {\bibinfo  {journal} {Phys. Rev. B}\ }\textbf {\bibinfo
  {volume} {103}},\ \bibinfo {pages} {L020403} (\bibinfo {year}
  {2021})}\BibitemShut {NoStop}%
\bibitem [{\citenamefont {Sampaio}\ \emph {et~al.}(2013)\citenamefont
  {Sampaio}, \citenamefont {Cros}, \citenamefont {Rohart}, \citenamefont
  {Thiaville},\ and\ \citenamefont {Fert}}]{Sampaio2013}%
  \BibitemOpen
  \bibfield  {author} {\bibinfo {author} {\bibfnamefont {J.}~\bibnamefont
  {Sampaio}}, \bibinfo {author} {\bibfnamefont {V.}~\bibnamefont {Cros}},
  \bibinfo {author} {\bibfnamefont {S.}~\bibnamefont {Rohart}}, \bibinfo
  {author} {\bibfnamefont {A.}~\bibnamefont {Thiaville}}, \ and\ \bibinfo
  {author} {\bibfnamefont {A.}~\bibnamefont {Fert}},\ }\href
  {http://dx.doi.org/10.1038/nnano.2013.210} {\bibfield  {journal} {\bibinfo
  {journal} {Nature Nanotechnology}\ }\textbf {\bibinfo {volume} {8}},\
  \bibinfo {pages} {839–844} (\bibinfo {year} {2013})}\BibitemShut {NoStop}%
\bibitem [{\citenamefont {Song}\ \emph {et~al.}(2020)\citenamefont {Song},
  \citenamefont {Jeong}, \citenamefont {Pan}, \citenamefont {Zhang},
  \citenamefont {Xia}, \citenamefont {Cha}, \citenamefont {Park}, \citenamefont
  {Kim}, \citenamefont {Finizio}, \citenamefont {Raabe}, \citenamefont {Chang},
  \citenamefont {Zhou}, \citenamefont {Zhao}, \citenamefont {Kang},
  \citenamefont {Ju},\ and\ \citenamefont {Woo}}]{Song2020}%
  \BibitemOpen
  \bibfield  {author} {\bibinfo {author} {\bibfnamefont {K.~M.}\ \bibnamefont
  {Song}}, \bibinfo {author} {\bibfnamefont {J.-S.}\ \bibnamefont {Jeong}},
  \bibinfo {author} {\bibfnamefont {B.}~\bibnamefont {Pan}}, \bibinfo {author}
  {\bibfnamefont {X.}~\bibnamefont {Zhang}}, \bibinfo {author} {\bibfnamefont
  {J.}~\bibnamefont {Xia}}, \bibinfo {author} {\bibfnamefont {S.}~\bibnamefont
  {Cha}}, \bibinfo {author} {\bibfnamefont {T.-E.}\ \bibnamefont {Park}},
  \bibinfo {author} {\bibfnamefont {K.}~\bibnamefont {Kim}}, \bibinfo {author}
  {\bibfnamefont {S.}~\bibnamefont {Finizio}}, \bibinfo {author} {\bibfnamefont
  {J.}~\bibnamefont {Raabe}}, \bibinfo {author} {\bibfnamefont
  {J.}~\bibnamefont {Chang}}, \bibinfo {author} {\bibfnamefont
  {Y.}~\bibnamefont {Zhou}}, \bibinfo {author} {\bibfnamefont {W.}~\bibnamefont
  {Zhao}}, \bibinfo {author} {\bibfnamefont {W.}~\bibnamefont {Kang}}, \bibinfo
  {author} {\bibfnamefont {H.}~\bibnamefont {Ju}}, \ and\ \bibinfo {author}
  {\bibfnamefont {S.}~\bibnamefont {Woo}},\ }\href {\doibase
  10.1038/s41928-020-0385-0} {\bibfield  {journal} {\bibinfo  {journal} {Nature
  Electronics}\ }\textbf {\bibinfo {volume} {3}},\ \bibinfo {pages} {148–155}
  (\bibinfo {year} {2020})}\BibitemShut {NoStop}%
\bibitem [{\citenamefont {Yokouchi}\ \emph {et~al.}(2022)\citenamefont
  {Yokouchi}, \citenamefont {Sugimoto}, \citenamefont {Rana}, \citenamefont
  {Seki}, \citenamefont {Ogawa}, \citenamefont {Shiomi}, \citenamefont
  {Kasai},\ and\ \citenamefont {Otani}}]{Yokouchi2022}%
  \BibitemOpen
  \bibfield  {author} {\bibinfo {author} {\bibfnamefont {T.}~\bibnamefont
  {Yokouchi}}, \bibinfo {author} {\bibfnamefont {S.}~\bibnamefont {Sugimoto}},
  \bibinfo {author} {\bibfnamefont {B.}~\bibnamefont {Rana}}, \bibinfo {author}
  {\bibfnamefont {S.}~\bibnamefont {Seki}}, \bibinfo {author} {\bibfnamefont
  {N.}~\bibnamefont {Ogawa}}, \bibinfo {author} {\bibfnamefont
  {Y.}~\bibnamefont {Shiomi}}, \bibinfo {author} {\bibfnamefont
  {S.}~\bibnamefont {Kasai}}, \ and\ \bibinfo {author} {\bibfnamefont
  {Y.}~\bibnamefont {Otani}},\ }\href
  {http://dx.doi.org/10.1126/sciadv.abq5652} {\bibfield  {journal} {\bibinfo
  {journal} {Science Advances}\ }\textbf {\bibinfo {volume} {8}} (\bibinfo
  {year} {2022})}\BibitemShut {NoStop}%
\bibitem [{\citenamefont {M\"uhlbauer}\ \emph {et~al.}(2009)\citenamefont
  {M\"uhlbauer}, \citenamefont {Binz}, \citenamefont {Jonietz}, \citenamefont
  {Pfleiderer}, \citenamefont {Rosch}, \citenamefont {Neubauer}, \citenamefont
  {Georgii},\ and\ \citenamefont {B\"oni}}]{Muhlbauer2009}%
  \BibitemOpen
  \bibfield  {author} {\bibinfo {author} {\bibfnamefont {S.}~\bibnamefont
  {M\"uhlbauer}}, \bibinfo {author} {\bibfnamefont {B.}~\bibnamefont {Binz}},
  \bibinfo {author} {\bibfnamefont {F.}~\bibnamefont {Jonietz}}, \bibinfo
  {author} {\bibfnamefont {C.}~\bibnamefont {Pfleiderer}}, \bibinfo {author}
  {\bibfnamefont {A.}~\bibnamefont {Rosch}}, \bibinfo {author} {\bibfnamefont
  {A.}~\bibnamefont {Neubauer}}, \bibinfo {author} {\bibfnamefont
  {R.}~\bibnamefont {Georgii}}, \ and\ \bibinfo {author} {\bibfnamefont
  {P.}~\bibnamefont {B\"oni}},\ }\href {\doibase 10.1126/science.1166767}
  {\bibfield  {journal} {\bibinfo  {journal} {Science}\ }\textbf {\bibinfo
  {volume} {323}},\ \bibinfo {pages} {915–919} (\bibinfo {year}
  {2009})}\BibitemShut {NoStop}%
\bibitem [{\citenamefont {Yu}\ \emph {et~al.}(2010)\citenamefont {Yu},
  \citenamefont {Onose}, \citenamefont {Kanazawa}, \citenamefont {Park},
  \citenamefont {Han}, \citenamefont {Matsui}, \citenamefont {Nagaosa},\ and\
  \citenamefont {Tokura}}]{Yu2010}%
  \BibitemOpen
  \bibfield  {author} {\bibinfo {author} {\bibfnamefont {X.~Z.}\ \bibnamefont
  {Yu}}, \bibinfo {author} {\bibfnamefont {Y.}~\bibnamefont {Onose}}, \bibinfo
  {author} {\bibfnamefont {N.}~\bibnamefont {Kanazawa}}, \bibinfo {author}
  {\bibfnamefont {J.~H.}\ \bibnamefont {Park}}, \bibinfo {author}
  {\bibfnamefont {J.~H.}\ \bibnamefont {Han}}, \bibinfo {author} {\bibfnamefont
  {Y.}~\bibnamefont {Matsui}}, \bibinfo {author} {\bibfnamefont
  {N.}~\bibnamefont {Nagaosa}}, \ and\ \bibinfo {author} {\bibfnamefont
  {Y.}~\bibnamefont {Tokura}},\ }\href {\doibase 10.1038/nature09124}
  {\bibfield  {journal} {\bibinfo  {journal} {Nature}\ }\textbf {\bibinfo
  {volume} {465}},\ \bibinfo {pages} {901–904} (\bibinfo {year}
  {2010})}\BibitemShut {NoStop}%
\bibitem [{\citenamefont {Adams}\ \emph {et~al.}(2011)\citenamefont {Adams},
  \citenamefont {M\"uhlbauer}, \citenamefont {Pfleiderer}, \citenamefont
  {Jonietz}, \citenamefont {Bauer}, \citenamefont {Neubauer}, \citenamefont
  {Georgii}, \citenamefont {B\"oni}, \citenamefont {Keiderling}, \citenamefont
  {Everschor}, \citenamefont {Garst},\ and\ \citenamefont
  {Rosch}}]{AdamsPRL2011}%
  \BibitemOpen
  \bibfield  {author} {\bibinfo {author} {\bibfnamefont {T.}~\bibnamefont
  {Adams}}, \bibinfo {author} {\bibfnamefont {S.}~\bibnamefont {M\"uhlbauer}},
  \bibinfo {author} {\bibfnamefont {C.}~\bibnamefont {Pfleiderer}}, \bibinfo
  {author} {\bibfnamefont {F.}~\bibnamefont {Jonietz}}, \bibinfo {author}
  {\bibfnamefont {A.}~\bibnamefont {Bauer}}, \bibinfo {author} {\bibfnamefont
  {A.}~\bibnamefont {Neubauer}}, \bibinfo {author} {\bibfnamefont
  {R.}~\bibnamefont {Georgii}}, \bibinfo {author} {\bibfnamefont
  {P.}~\bibnamefont {B\"oni}}, \bibinfo {author} {\bibfnamefont
  {U.}~\bibnamefont {Keiderling}}, \bibinfo {author} {\bibfnamefont
  {K.}~\bibnamefont {Everschor}}, \bibinfo {author} {\bibfnamefont
  {M.}~\bibnamefont {Garst}}, \ and\ \bibinfo {author} {\bibfnamefont
  {A.}~\bibnamefont {Rosch}},\ }\href {\doibase 10.1103/PhysRevLett.107.217206}
  {\bibfield  {journal} {\bibinfo  {journal} {Phys. Rev. Lett.}\ }\textbf
  {\bibinfo {volume} {107}},\ \bibinfo {pages} {217206} (\bibinfo {year}
  {2011})}\BibitemShut {NoStop}%
\bibitem [{\citenamefont {Rosales}\ \emph {et~al.}(2015)\citenamefont
  {Rosales}, \citenamefont {Cabra},\ and\ \citenamefont
  {Pujol}}]{rosales2015three}%
  \BibitemOpen
  \bibfield  {author} {\bibinfo {author} {\bibfnamefont {H.~D.}\ \bibnamefont
  {Rosales}}, \bibinfo {author} {\bibfnamefont {D.~C.}\ \bibnamefont {Cabra}},
  \ and\ \bibinfo {author} {\bibfnamefont {P.}~\bibnamefont {Pujol}},\ }\href
  {\doibase https://doi.org/10.1103/PhysRevB.92.214439} {\bibfield  {journal}
  {\bibinfo  {journal} {Phys. Rev. B}\ }\textbf {\bibinfo {volume} {92}},\
  \bibinfo {pages} {214439} (\bibinfo {year} {2015})}\BibitemShut {NoStop}%
\bibitem [{\citenamefont {Gao}\ \emph {et~al.}(2020)\citenamefont {Gao},
  \citenamefont {Rosales}, \citenamefont {G{\'o}mez~Albarrac{\'\i}n},
  \citenamefont {Tsurkan}, \citenamefont {Kaur}, \citenamefont {Fennell},
  \citenamefont {Steffens}, \citenamefont {Boehm}, \citenamefont
  {{\v{C}}erm{\'a}k}, \citenamefont {Schneidewind} \emph
  {et~al.}}]{gao2020fractional}%
  \BibitemOpen
  \bibfield  {author} {\bibinfo {author} {\bibfnamefont {S.}~\bibnamefont
  {Gao}}, \bibinfo {author} {\bibfnamefont {H.~D.}\ \bibnamefont {Rosales}},
  \bibinfo {author} {\bibfnamefont {F.~A.}\ \bibnamefont
  {G{\'o}mez~Albarrac{\'\i}n}}, \bibinfo {author} {\bibfnamefont
  {V.}~\bibnamefont {Tsurkan}}, \bibinfo {author} {\bibfnamefont
  {G.}~\bibnamefont {Kaur}}, \bibinfo {author} {\bibfnamefont {T.}~\bibnamefont
  {Fennell}}, \bibinfo {author} {\bibfnamefont {P.}~\bibnamefont {Steffens}},
  \bibinfo {author} {\bibfnamefont {M.}~\bibnamefont {Boehm}}, \bibinfo
  {author} {\bibfnamefont {P.}~\bibnamefont {{\v{C}}erm{\'a}k}}, \bibinfo
  {author} {\bibfnamefont {A.}~\bibnamefont {Schneidewind}},  \emph {et~al.},\
  }\href {\doibase https://doi.org/10.1038/s41586-020-2716} {\bibfield
  {journal} {\bibinfo  {journal} {Nature}\ }\textbf {\bibinfo {volume} {586}},\
  \bibinfo {pages} {37} (\bibinfo {year} {2020})}\BibitemShut {NoStop}%
\bibitem [{\citenamefont {Mohylna}\ \emph {et~al.}(2022)\citenamefont
  {Mohylna}, \citenamefont {G{\'o}mez~Albarrac{\'\i}n}, \citenamefont
  {{\v{Z}}ukovi{\v{c}}},\ and\ \citenamefont
  {Rosales}}]{mohylna2022spontaneous}%
  \BibitemOpen
  \bibfield  {author} {\bibinfo {author} {\bibfnamefont {M.}~\bibnamefont
  {Mohylna}}, \bibinfo {author} {\bibfnamefont {F.~A.}\ \bibnamefont
  {G{\'o}mez~Albarrac{\'\i}n}}, \bibinfo {author} {\bibfnamefont
  {M.}~\bibnamefont {{\v{Z}}ukovi{\v{c}}}}, \ and\ \bibinfo {author}
  {\bibfnamefont {H.~D.}\ \bibnamefont {Rosales}},\ }\href {\doibase
  https://doi.org/10.1103/PhysRevB.106.224406} {\bibfield  {journal} {\bibinfo
  {journal} {Phys. Rev. B}\ }\textbf {\bibinfo {volume} {106}},\ \bibinfo
  {pages} {224406} (\bibinfo {year} {2022})}\BibitemShut {NoStop}%
\bibitem [{\citenamefont {Rosales}\ \emph {et~al.}(2023)\citenamefont
  {Rosales}, \citenamefont {G{\'o}mez~Albarrac{\'\i}n}, \citenamefont {Pujol},\
  and\ \citenamefont {Jaubert}}]{rosales2023skyrmion}%
  \BibitemOpen
  \bibfield  {author} {\bibinfo {author} {\bibfnamefont {H.~D.}\ \bibnamefont
  {Rosales}}, \bibinfo {author} {\bibfnamefont {F.~A.}\ \bibnamefont
  {G{\'o}mez~Albarrac{\'\i}n}}, \bibinfo {author} {\bibfnamefont
  {P.}~\bibnamefont {Pujol}}, \ and\ \bibinfo {author} {\bibfnamefont {L.~D.}\
  \bibnamefont {Jaubert}},\ }\href {\doibase
  https://doi.org/10.1103/PhysRevLett.130.106703} {\bibfield  {journal}
  {\bibinfo  {journal} {Phys. Rev. Lett.}\ }\textbf {\bibinfo {volume} {130}},\
  \bibinfo {pages} {106703} (\bibinfo {year} {2023})}\BibitemShut {NoStop}%
\bibitem [{\citenamefont {G{\'o}mez~Albarrac{\'\i}n}\ \emph
  {et~al.}(2024)\citenamefont {G{\'o}mez~Albarrac{\'\i}n}, \citenamefont
  {Rosales}, \citenamefont {Udagawa}, \citenamefont {Pujol},\ and\
  \citenamefont {Jaubert}}]{gomez2024chiral}%
  \BibitemOpen
  \bibfield  {author} {\bibinfo {author} {\bibfnamefont {F.~A.}\ \bibnamefont
  {G{\'o}mez~Albarrac{\'\i}n}}, \bibinfo {author} {\bibfnamefont {H.~D.}\
  \bibnamefont {Rosales}}, \bibinfo {author} {\bibfnamefont {M.}~\bibnamefont
  {Udagawa}}, \bibinfo {author} {\bibfnamefont {P.}~\bibnamefont {Pujol}}, \
  and\ \bibinfo {author} {\bibfnamefont {L.~D.}\ \bibnamefont {Jaubert}},\
  }\href {\doibase https://doi.org/10.1103/PhysRevB.109.064426} {\bibfield
  {journal} {\bibinfo  {journal} {Phys. Rev. B}\ }\textbf {\bibinfo {volume}
  {109}},\ \bibinfo {pages} {064426} (\bibinfo {year} {2024})}\BibitemShut
  {NoStop}%
\bibitem [{\citenamefont {Hoshino}\ and\ \citenamefont
  {Nagaosa}(2018)}]{HoshinoPRB2018}%
  \BibitemOpen
  \bibfield  {author} {\bibinfo {author} {\bibfnamefont {S.}~\bibnamefont
  {Hoshino}}\ and\ \bibinfo {author} {\bibfnamefont {N.}~\bibnamefont
  {Nagaosa}},\ }\href {\doibase 10.1103/PhysRevB.97.024413} {\bibfield
  {journal} {\bibinfo  {journal} {Phys. Rev. B}\ }\textbf {\bibinfo {volume}
  {97}},\ \bibinfo {pages} {024413} (\bibinfo {year} {2018})}\BibitemShut
  {NoStop}%
\bibitem [{\citenamefont {Chudnovsky}\ and\ \citenamefont
  {Garanin}(2018)}]{Chudnovsky2018}%
  \BibitemOpen
  \bibfield  {author} {\bibinfo {author} {\bibfnamefont {E.~M.}\ \bibnamefont
  {Chudnovsky}}\ and\ \bibinfo {author} {\bibfnamefont {D.~A.}\ \bibnamefont
  {Garanin}},\ }\href {\doibase 10.1088/1367-2630/aab576} {\bibfield  {journal}
  {\bibinfo  {journal} {New Journal of Physics}\ }\textbf {\bibinfo {volume}
  {20}},\ \bibinfo {pages} {033006} (\bibinfo {year} {2018})}\BibitemShut
  {NoStop}%
\bibitem [{\citenamefont {Koshibae}\ and\ \citenamefont
  {Nagaosa}(2018)}]{Koshibae2018}%
  \BibitemOpen
  \bibfield  {author} {\bibinfo {author} {\bibfnamefont {W.}~\bibnamefont
  {Koshibae}}\ and\ \bibinfo {author} {\bibfnamefont {N.}~\bibnamefont
  {Nagaosa}},\ }\href {http://dx.doi.org/10.1038/s41598-018-24693-5} {\bibfield
   {journal} {\bibinfo  {journal} {Scientific Reports}\ }\textbf {\bibinfo
  {volume} {8}} (\bibinfo {year} {2018})}\BibitemShut {NoStop}%
\bibitem [{\citenamefont {Silva}\ \emph {et~al.}(2014)\citenamefont {Silva},
  \citenamefont {Secchin}, \citenamefont {Moura-Melo}, \citenamefont
  {Pereira},\ and\ \citenamefont {Stamps}}]{SilvaPRB2014}%
  \BibitemOpen
  \bibfield  {author} {\bibinfo {author} {\bibfnamefont {R.~L.}\ \bibnamefont
  {Silva}}, \bibinfo {author} {\bibfnamefont {L.~D.}\ \bibnamefont {Secchin}},
  \bibinfo {author} {\bibfnamefont {W.~A.}\ \bibnamefont {Moura-Melo}},
  \bibinfo {author} {\bibfnamefont {A.~R.}\ \bibnamefont {Pereira}}, \ and\
  \bibinfo {author} {\bibfnamefont {R.~L.}\ \bibnamefont {Stamps}},\ }\href
  {\doibase 10.1103/PhysRevB.89.054434} {\bibfield  {journal} {\bibinfo
  {journal} {Phys. Rev. B}\ }\textbf {\bibinfo {volume} {89}},\ \bibinfo
  {pages} {054434} (\bibinfo {year} {2014})}\BibitemShut {NoStop}%
\bibitem [{\citenamefont {D\'{\i}az}\ \emph {et~al.}(2017)\citenamefont
  {D\'{\i}az}, \citenamefont {Reichhardt}, \citenamefont {Arovas},
  \citenamefont {Saxena},\ and\ \citenamefont {Reichhardt}}]{DiazPRB2017}%
  \BibitemOpen
  \bibfield  {author} {\bibinfo {author} {\bibfnamefont {S.~A.}\ \bibnamefont
  {D\'{\i}az}}, \bibinfo {author} {\bibfnamefont {C.~J.~O.}\ \bibnamefont
  {Reichhardt}}, \bibinfo {author} {\bibfnamefont {D.~P.}\ \bibnamefont
  {Arovas}}, \bibinfo {author} {\bibfnamefont {A.}~\bibnamefont {Saxena}}, \
  and\ \bibinfo {author} {\bibfnamefont {C.}~\bibnamefont {Reichhardt}},\
  }\href {\doibase 10.1103/PhysRevB.96.085106} {\bibfield  {journal} {\bibinfo
  {journal} {Phys. Rev. B}\ }\textbf {\bibinfo {volume} {96}},\ \bibinfo
  {pages} {085106} (\bibinfo {year} {2017})}\BibitemShut {NoStop}%
\bibitem [{\citenamefont {Karube}\ \emph {et~al.}(2018)\citenamefont {Karube},
  \citenamefont {White}, \citenamefont {Morikawa}, \citenamefont {Dewhurst},
  \citenamefont {Cubitt}, \citenamefont {Kikkawa}, \citenamefont {Yu},
  \citenamefont {Tokunaga}, \citenamefont {Arima}, \citenamefont {Rønnow},
  \citenamefont {Tokura},\ and\ \citenamefont {Taguchi}}]{Karube2018}%
  \BibitemOpen
  \bibfield  {author} {\bibinfo {author} {\bibfnamefont {K.}~\bibnamefont
  {Karube}}, \bibinfo {author} {\bibfnamefont {J.~S.}\ \bibnamefont {White}},
  \bibinfo {author} {\bibfnamefont {D.}~\bibnamefont {Morikawa}}, \bibinfo
  {author} {\bibfnamefont {C.~D.}\ \bibnamefont {Dewhurst}}, \bibinfo {author}
  {\bibfnamefont {R.}~\bibnamefont {Cubitt}}, \bibinfo {author} {\bibfnamefont
  {A.}~\bibnamefont {Kikkawa}}, \bibinfo {author} {\bibfnamefont
  {X.}~\bibnamefont {Yu}}, \bibinfo {author} {\bibfnamefont {Y.}~\bibnamefont
  {Tokunaga}}, \bibinfo {author} {\bibfnamefont {T.-h.}\ \bibnamefont {Arima}},
  \bibinfo {author} {\bibfnamefont {H.~M.}\ \bibnamefont {Rønnow}}, \bibinfo
  {author} {\bibfnamefont {Y.}~\bibnamefont {Tokura}}, \ and\ \bibinfo {author}
  {\bibfnamefont {Y.}~\bibnamefont {Taguchi}},\ }\href
  {http://dx.doi.org/10.1126/sciadv.aar7043} {\bibfield  {journal} {\bibinfo
  {journal} {Science Advances}\ }\textbf {\bibinfo {volume} {4}} (\bibinfo
  {year} {2018})}\BibitemShut {NoStop}%
\bibitem [{\citenamefont {Reichhardt}\ \emph {et~al.}(2022)\citenamefont
  {Reichhardt}, \citenamefont {Reichhardt},\ and\ \citenamefont {Milo\ifmmode
  \check{s}\else \v{s}\fi{}evi\ifmmode~\acute{c}\else
  \'{c}\fi{}}}]{Reichhardt-RMP2022}%
  \BibitemOpen
  \bibfield  {author} {\bibinfo {author} {\bibfnamefont {C.}~\bibnamefont
  {Reichhardt}}, \bibinfo {author} {\bibfnamefont {C.~J.~O.}\ \bibnamefont
  {Reichhardt}}, \ and\ \bibinfo {author} {\bibfnamefont {M.~V.}\ \bibnamefont
  {Milo\ifmmode \check{s}\else \v{s}\fi{}evi\ifmmode~\acute{c}\else
  \'{c}\fi{}}},\ }\href {https://link.aps.org/doi/10.1103/RevModPhys.94.035005}
  {\bibfield  {journal} {\bibinfo  {journal} {Rev. Mod. Phys.}\ }\textbf
  {\bibinfo {volume} {94}},\ \bibinfo {pages} {035005} (\bibinfo {year}
  {2022})}\BibitemShut {NoStop}%
\bibitem [{\citenamefont {Huang}\ \emph {et~al.}(2020)\citenamefont {Huang},
  \citenamefont {Sch\"{o}nenberger}, \citenamefont {Cantoni}, \citenamefont
  {Heinen}, \citenamefont {Magrez}, \citenamefont {Rosch}, \citenamefont
  {Carbone},\ and\ \citenamefont {Rønnow}}]{Huang2020}%
  \BibitemOpen
  \bibfield  {author} {\bibinfo {author} {\bibfnamefont {P.}~\bibnamefont
  {Huang}}, \bibinfo {author} {\bibfnamefont {T.}~\bibnamefont
  {Sch\"{o}nenberger}}, \bibinfo {author} {\bibfnamefont {M.}~\bibnamefont
  {Cantoni}}, \bibinfo {author} {\bibfnamefont {L.}~\bibnamefont {Heinen}},
  \bibinfo {author} {\bibfnamefont {A.}~\bibnamefont {Magrez}}, \bibinfo
  {author} {\bibfnamefont {A.}~\bibnamefont {Rosch}}, \bibinfo {author}
  {\bibfnamefont {F.}~\bibnamefont {Carbone}}, \ and\ \bibinfo {author}
  {\bibfnamefont {H.~M.}\ \bibnamefont {Rønnow}},\ }\href {\doibase
  10.1038/s41565-020-0716-3} {\bibfield  {journal} {\bibinfo  {journal} {Nat.
  Nanotechnol.}\ }\textbf {\bibinfo {volume} {15}},\ \bibinfo {pages}
  {761–767} (\bibinfo {year} {2020})}\BibitemShut {NoStop}%
\bibitem [{\citenamefont {Esposito}\ \emph {et~al.}(2020)\citenamefont
  {Esposito}, \citenamefont {Zheng}, \citenamefont {Seaberg}, \citenamefont
  {Montoya}, \citenamefont {Holladay}, \citenamefont {Reid}, \citenamefont
  {Streubel}, \citenamefont {Lee}, \citenamefont {Shen}, \citenamefont
  {Koralek}, \citenamefont {Coslovich}, \citenamefont {Walter}, \citenamefont
  {Zohar}, \citenamefont {Thampy}, \citenamefont {Lin}, \citenamefont {Hart},
  \citenamefont {Nakahara}, \citenamefont {Fischer}, \citenamefont {Colocho},
  \citenamefont {Lutman}, \citenamefont {Decker}, \citenamefont {Sinha},
  \citenamefont {Fullerton}, \citenamefont {Kevan}, \citenamefont {Roy},
  \citenamefont {Dunne},\ and\ \citenamefont {Turner}}]{Esposito2020}%
  \BibitemOpen
  \bibfield  {author} {\bibinfo {author} {\bibfnamefont {V.}~\bibnamefont
  {Esposito}}, \bibinfo {author} {\bibfnamefont {X.~Y.}\ \bibnamefont {Zheng}},
  \bibinfo {author} {\bibfnamefont {M.~H.}\ \bibnamefont {Seaberg}}, \bibinfo
  {author} {\bibfnamefont {S.~A.}\ \bibnamefont {Montoya}}, \bibinfo {author}
  {\bibfnamefont {B.}~\bibnamefont {Holladay}}, \bibinfo {author}
  {\bibfnamefont {A.~H.}\ \bibnamefont {Reid}}, \bibinfo {author}
  {\bibfnamefont {R.}~\bibnamefont {Streubel}}, \bibinfo {author}
  {\bibfnamefont {J.~C.~T.}\ \bibnamefont {Lee}}, \bibinfo {author}
  {\bibfnamefont {L.}~\bibnamefont {Shen}}, \bibinfo {author} {\bibfnamefont
  {J.~D.}\ \bibnamefont {Koralek}}, \bibinfo {author} {\bibfnamefont
  {G.}~\bibnamefont {Coslovich}}, \bibinfo {author} {\bibfnamefont
  {P.}~\bibnamefont {Walter}}, \bibinfo {author} {\bibfnamefont
  {S.}~\bibnamefont {Zohar}}, \bibinfo {author} {\bibfnamefont
  {V.}~\bibnamefont {Thampy}}, \bibinfo {author} {\bibfnamefont {M.~F.}\
  \bibnamefont {Lin}}, \bibinfo {author} {\bibfnamefont {P.}~\bibnamefont
  {Hart}}, \bibinfo {author} {\bibfnamefont {K.}~\bibnamefont {Nakahara}},
  \bibinfo {author} {\bibfnamefont {P.}~\bibnamefont {Fischer}}, \bibinfo
  {author} {\bibfnamefont {W.}~\bibnamefont {Colocho}}, \bibinfo {author}
  {\bibfnamefont {A.}~\bibnamefont {Lutman}}, \bibinfo {author} {\bibfnamefont
  {F.-J.}\ \bibnamefont {Decker}}, \bibinfo {author} {\bibfnamefont {S.~K.}\
  \bibnamefont {Sinha}}, \bibinfo {author} {\bibfnamefont {E.~E.}\ \bibnamefont
  {Fullerton}}, \bibinfo {author} {\bibfnamefont {S.~D.}\ \bibnamefont
  {Kevan}}, \bibinfo {author} {\bibfnamefont {S.}~\bibnamefont {Roy}}, \bibinfo
  {author} {\bibfnamefont {M.}~\bibnamefont {Dunne}}, \ and\ \bibinfo {author}
  {\bibfnamefont {J.~J.}\ \bibnamefont {Turner}},\ }\href
  {http://dx.doi.org/10.1063/5.0004879} {\bibfield  {journal} {\bibinfo
  {journal} {Applied Physics Letters}\ }\textbf {\bibinfo {volume} {116}}
  (\bibinfo {year} {2020})}\BibitemShut {NoStop}%
\bibitem [{\citenamefont {Mohanta}\ \emph {et~al.}(2020)\citenamefont
  {Mohanta}, \citenamefont {Christianson}, \citenamefont {Okamoto},\ and\
  \citenamefont {Dagotto}}]{Mohanta2020}%
  \BibitemOpen
  \bibfield  {author} {\bibinfo {author} {\bibfnamefont {N.}~\bibnamefont
  {Mohanta}}, \bibinfo {author} {\bibfnamefont {A.~D.}\ \bibnamefont
  {Christianson}}, \bibinfo {author} {\bibfnamefont {S.}~\bibnamefont
  {Okamoto}}, \ and\ \bibinfo {author} {\bibfnamefont {E.}~\bibnamefont
  {Dagotto}},\ }\href {http://dx.doi.org/10.1038/s42005-020-00489-w} {\bibfield
   {journal} {\bibinfo  {journal} {Communications Physics}\ }\textbf {\bibinfo
  {volume} {3}} (\bibinfo {year} {2020})}\BibitemShut {NoStop}%
\bibitem [{\citenamefont {Karube}\ \emph {et~al.}(2020)\citenamefont {Karube},
  \citenamefont {White}, \citenamefont {Ukleev}, \citenamefont {Dewhurst},
  \citenamefont {Cubitt}, \citenamefont {Kikkawa}, \citenamefont {Tokunaga},
  \citenamefont {R\o{}nnow}, \citenamefont {Tokura},\ and\ \citenamefont
  {Taguchi}}]{KarubePRB2020}%
  \BibitemOpen
  \bibfield  {author} {\bibinfo {author} {\bibfnamefont {K.}~\bibnamefont
  {Karube}}, \bibinfo {author} {\bibfnamefont {J.~S.}\ \bibnamefont {White}},
  \bibinfo {author} {\bibfnamefont {V.}~\bibnamefont {Ukleev}}, \bibinfo
  {author} {\bibfnamefont {C.~D.}\ \bibnamefont {Dewhurst}}, \bibinfo {author}
  {\bibfnamefont {R.}~\bibnamefont {Cubitt}}, \bibinfo {author} {\bibfnamefont
  {A.}~\bibnamefont {Kikkawa}}, \bibinfo {author} {\bibfnamefont
  {Y.}~\bibnamefont {Tokunaga}}, \bibinfo {author} {\bibfnamefont {H.~M.}\
  \bibnamefont {R\o{}nnow}}, \bibinfo {author} {\bibfnamefont {Y.}~\bibnamefont
  {Tokura}}, \ and\ \bibinfo {author} {\bibfnamefont {Y.}~\bibnamefont
  {Taguchi}},\ }\href {\doibase 10.1103/PhysRevB.102.064408} {\bibfield
  {journal} {\bibinfo  {journal} {Phys. Rev. B}\ }\textbf {\bibinfo {volume}
  {102}},\ \bibinfo {pages} {064408} (\bibinfo {year} {2020})}\BibitemShut
  {NoStop}%
\bibitem [{\citenamefont {Meisenheimer}\ \emph {et~al.}(2023)\citenamefont
  {Meisenheimer}, \citenamefont {Zhang}, \citenamefont {Raftrey}, \citenamefont
  {Chen}, \citenamefont {Shao}, \citenamefont {Chan}, \citenamefont {Yalisove},
  \citenamefont {Chen}, \citenamefont {Yao}, \citenamefont {Scott},
  \citenamefont {Wu}, \citenamefont {Muller}, \citenamefont {Fischer},
  \citenamefont {Birgeneau},\ and\ \citenamefont {Ramesh}}]{Meisenheimer2023}%
  \BibitemOpen
  \bibfield  {author} {\bibinfo {author} {\bibfnamefont {P.}~\bibnamefont
  {Meisenheimer}}, \bibinfo {author} {\bibfnamefont {H.}~\bibnamefont {Zhang}},
  \bibinfo {author} {\bibfnamefont {D.}~\bibnamefont {Raftrey}}, \bibinfo
  {author} {\bibfnamefont {X.}~\bibnamefont {Chen}}, \bibinfo {author}
  {\bibfnamefont {Y.-T.}\ \bibnamefont {Shao}}, \bibinfo {author}
  {\bibfnamefont {Y.-T.}\ \bibnamefont {Chan}}, \bibinfo {author}
  {\bibfnamefont {R.}~\bibnamefont {Yalisove}}, \bibinfo {author}
  {\bibfnamefont {R.}~\bibnamefont {Chen}}, \bibinfo {author} {\bibfnamefont
  {J.}~\bibnamefont {Yao}}, \bibinfo {author} {\bibfnamefont {M.~C.}\
  \bibnamefont {Scott}}, \bibinfo {author} {\bibfnamefont {W.}~\bibnamefont
  {Wu}}, \bibinfo {author} {\bibfnamefont {D.~A.}\ \bibnamefont {Muller}},
  \bibinfo {author} {\bibfnamefont {P.}~\bibnamefont {Fischer}}, \bibinfo
  {author} {\bibfnamefont {R.~J.}\ \bibnamefont {Birgeneau}}, \ and\ \bibinfo
  {author} {\bibfnamefont {R.}~\bibnamefont {Ramesh}},\ }\href
  {http://dx.doi.org/10.1038/s41467-023-39442-0} {\bibfield  {journal}
  {\bibinfo  {journal} {Nat. Commun.}\ }\textbf {\bibinfo {volume} {14}}
  (\bibinfo {year} {2023})}\BibitemShut {NoStop}%
\bibitem [{\citenamefont {Roldán-Molina}\ \emph {et~al.}(2016)\citenamefont
  {Roldán-Molina}, \citenamefont {Nunez},\ and\ \citenamefont
  {Fernández-Rossier}}]{roldanNJP2016}%
  \BibitemOpen
  \bibfield  {author} {\bibinfo {author} {\bibfnamefont {A.}~\bibnamefont
  {Roldán-Molina}}, \bibinfo {author} {\bibfnamefont {A.~S.}\ \bibnamefont
  {Nunez}}, \ and\ \bibinfo {author} {\bibfnamefont {J.}~\bibnamefont
  {Fernández-Rossier}},\ }\href {\doibase 10.1088/1367-2630/18/4/045015}
  {\bibfield  {journal} {\bibinfo  {journal} {New Journal of Physics}\ }\textbf
  {\bibinfo {volume} {18}},\ \bibinfo {pages} {045015} (\bibinfo {year}
  {2016})}\BibitemShut {NoStop}%
\bibitem [{\citenamefont {D\'{\i}az}\ \emph {et~al.}(2019)\citenamefont
  {D\'{\i}az}, \citenamefont {Klinovaja},\ and\ \citenamefont
  {Loss}}]{diazPRL2019}%
  \BibitemOpen
  \bibfield  {author} {\bibinfo {author} {\bibfnamefont {S.~A.}\ \bibnamefont
  {D\'{\i}az}}, \bibinfo {author} {\bibfnamefont {J.}~\bibnamefont
  {Klinovaja}}, \ and\ \bibinfo {author} {\bibfnamefont {D.}~\bibnamefont
  {Loss}},\ }\href {\doibase 10.1103/PhysRevLett.122.187203} {\bibfield
  {journal} {\bibinfo  {journal} {Phys. Rev. Lett.}\ }\textbf {\bibinfo
  {volume} {122}},\ \bibinfo {pages} {187203} (\bibinfo {year}
  {2019})}\BibitemShut {NoStop}%
\bibitem [{\citenamefont {D\'{\i}az}\ \emph {et~al.}(2020)\citenamefont
  {D\'{\i}az}, \citenamefont {Hirosawa}, \citenamefont {Klinovaja},\ and\
  \citenamefont {Loss}}]{diazPRB2020}%
  \BibitemOpen
  \bibfield  {author} {\bibinfo {author} {\bibfnamefont {S.~A.}\ \bibnamefont
  {D\'{\i}az}}, \bibinfo {author} {\bibfnamefont {T.}~\bibnamefont {Hirosawa}},
  \bibinfo {author} {\bibfnamefont {J.}~\bibnamefont {Klinovaja}}, \ and\
  \bibinfo {author} {\bibfnamefont {D.}~\bibnamefont {Loss}},\ }\href {\doibase
  10.1103/PhysRevResearch.2.013231} {\bibfield  {journal} {\bibinfo  {journal}
  {Phys. Rev. Res.}\ }\textbf {\bibinfo {volume} {2}},\ \bibinfo {pages}
  {013231} (\bibinfo {year} {2020})}\BibitemShut {NoStop}%
\bibitem [{\citenamefont {G{\'o}mez~Albarrac{\'\i}n}\ \emph
  {et~al.}(2021)\citenamefont {G{\'o}mez~Albarrac{\'\i}n}, \citenamefont
  {Rosales},\ and\ \citenamefont {Pujol}}]{albarracin2021chiral}%
  \BibitemOpen
  \bibfield  {author} {\bibinfo {author} {\bibfnamefont {F.}~\bibnamefont
  {G{\'o}mez~Albarrac{\'\i}n}}, \bibinfo {author} {\bibfnamefont {H.~D.}\
  \bibnamefont {Rosales}}, \ and\ \bibinfo {author} {\bibfnamefont
  {P.}~\bibnamefont {Pujol}},\ }\href {\doibase
  https://doi.org/10.1103/PhysRevB.103.054405} {\bibfield  {journal} {\bibinfo
  {journal} {Phys. Rev. B}\ }\textbf {\bibinfo {volume} {103}},\ \bibinfo
  {pages} {054405} (\bibinfo {year} {2021})}\BibitemShut {NoStop}%
\bibitem [{\citenamefont {Akazawa}\ \emph {et~al.}(2022)\citenamefont
  {Akazawa}, \citenamefont {Lee}, \citenamefont {Takeda}, \citenamefont
  {Fujima}, \citenamefont {Tokunaga}, \citenamefont {Arima}, \citenamefont
  {Han},\ and\ \citenamefont {Yamashita}}]{AkazawaPRR2022}%
  \BibitemOpen
  \bibfield  {author} {\bibinfo {author} {\bibfnamefont {M.}~\bibnamefont
  {Akazawa}}, \bibinfo {author} {\bibfnamefont {H.-Y.}\ \bibnamefont {Lee}},
  \bibinfo {author} {\bibfnamefont {H.}~\bibnamefont {Takeda}}, \bibinfo
  {author} {\bibfnamefont {Y.}~\bibnamefont {Fujima}}, \bibinfo {author}
  {\bibfnamefont {Y.}~\bibnamefont {Tokunaga}}, \bibinfo {author}
  {\bibfnamefont {T.-h.}\ \bibnamefont {Arima}}, \bibinfo {author}
  {\bibfnamefont {J.~H.}\ \bibnamefont {Han}}, \ and\ \bibinfo {author}
  {\bibfnamefont {M.}~\bibnamefont {Yamashita}},\ }\href {\doibase
  10.1103/PhysRevResearch.4.043085} {\bibfield  {journal} {\bibinfo  {journal}
  {Phys. Rev. Res.}\ }\textbf {\bibinfo {volume} {4}},\ \bibinfo {pages}
  {043085} (\bibinfo {year} {2022})}\BibitemShut {NoStop}%
\bibitem [{\citenamefont {Weber}\ \emph {et~al.}(2022)\citenamefont {Weber},
  \citenamefont {Fobes}, \citenamefont {Waizner}, \citenamefont {Steffens},
  \citenamefont {Tucker}, \citenamefont {B\"{o}hm}, \citenamefont {Beddrich},
  \citenamefont {Franz}, \citenamefont {Gabold}, \citenamefont {Bewley},
  \citenamefont {Voneshen}, \citenamefont {Skoulatos}, \citenamefont {Georgii},
  \citenamefont {Ehlers}, \citenamefont {Bauer}, \citenamefont {Pfleiderer},
  \citenamefont {B\"{o}ni}, \citenamefont {Janoschek},\ and\ \citenamefont
  {Garst}}]{Weber2022}%
  \BibitemOpen
  \bibfield  {author} {\bibinfo {author} {\bibfnamefont {T.}~\bibnamefont
  {Weber}}, \bibinfo {author} {\bibfnamefont {D.~M.}\ \bibnamefont {Fobes}},
  \bibinfo {author} {\bibfnamefont {J.}~\bibnamefont {Waizner}}, \bibinfo
  {author} {\bibfnamefont {P.}~\bibnamefont {Steffens}}, \bibinfo {author}
  {\bibfnamefont {G.~S.}\ \bibnamefont {Tucker}}, \bibinfo {author}
  {\bibfnamefont {M.}~\bibnamefont {B\"{o}hm}}, \bibinfo {author}
  {\bibfnamefont {L.}~\bibnamefont {Beddrich}}, \bibinfo {author}
  {\bibfnamefont {C.}~\bibnamefont {Franz}}, \bibinfo {author} {\bibfnamefont
  {H.}~\bibnamefont {Gabold}}, \bibinfo {author} {\bibfnamefont
  {R.}~\bibnamefont {Bewley}}, \bibinfo {author} {\bibfnamefont
  {D.}~\bibnamefont {Voneshen}}, \bibinfo {author} {\bibfnamefont
  {M.}~\bibnamefont {Skoulatos}}, \bibinfo {author} {\bibfnamefont
  {R.}~\bibnamefont {Georgii}}, \bibinfo {author} {\bibfnamefont
  {G.}~\bibnamefont {Ehlers}}, \bibinfo {author} {\bibfnamefont
  {A.}~\bibnamefont {Bauer}}, \bibinfo {author} {\bibfnamefont
  {C.}~\bibnamefont {Pfleiderer}}, \bibinfo {author} {\bibfnamefont
  {P.}~\bibnamefont {B\"{o}ni}}, \bibinfo {author} {\bibfnamefont
  {M.}~\bibnamefont {Janoschek}}, \ and\ \bibinfo {author} {\bibfnamefont
  {M.}~\bibnamefont {Garst}},\ }\href {\doibase 10.1126/science.abe4441}
  {\bibfield  {journal} {\bibinfo  {journal} {Science}\ }\textbf {\bibinfo
  {volume} {375}},\ \bibinfo {pages} {1025–1030} (\bibinfo {year}
  {2022})}\BibitemShut {NoStop}%
\bibitem [{\citenamefont {Hirosawa}\ \emph {et~al.}(2020)\citenamefont
  {Hirosawa}, \citenamefont {D\'{\i}az}, \citenamefont {Klinovaja},\ and\
  \citenamefont {Loss}}]{HirosawaPRL2020}%
  \BibitemOpen
  \bibfield  {author} {\bibinfo {author} {\bibfnamefont {T.}~\bibnamefont
  {Hirosawa}}, \bibinfo {author} {\bibfnamefont {S.~A.}\ \bibnamefont
  {D\'{\i}az}}, \bibinfo {author} {\bibfnamefont {J.}~\bibnamefont
  {Klinovaja}}, \ and\ \bibinfo {author} {\bibfnamefont {D.}~\bibnamefont
  {Loss}},\ }\href {\doibase 10.1103/PhysRevLett.125.207204} {\bibfield
  {journal} {\bibinfo  {journal} {Phys. Rev. Lett.}\ }\textbf {\bibinfo
  {volume} {125}},\ \bibinfo {pages} {207204} (\bibinfo {year}
  {2020})}\BibitemShut {NoStop}%
\bibitem [{\citenamefont {Zhuo}\ \emph {et~al.}(2023)\citenamefont {Zhuo},
  \citenamefont {Kang}, \citenamefont {Manchon},\ and\ \citenamefont
  {Cheng}}]{Zhuo2023}%
  \BibitemOpen
  \bibfield  {author} {\bibinfo {author} {\bibfnamefont {F.}~\bibnamefont
  {Zhuo}}, \bibinfo {author} {\bibfnamefont {J.}~\bibnamefont {Kang}}, \bibinfo
  {author} {\bibfnamefont {A.}~\bibnamefont {Manchon}}, \ and\ \bibinfo
  {author} {\bibfnamefont {Z.}~\bibnamefont {Cheng}},\ }\href
  {http://dx.doi.org/10.1002/apxr.202300054} {\bibfield  {journal} {\bibinfo
  {journal} {Advanced Physics Research}\ } (\bibinfo {year}
  {2023})}\BibitemShut {NoStop}%
\bibitem [{\citenamefont {Onose}\ \emph {et~al.}(2010)\citenamefont {Onose},
  \citenamefont {Ideue}, \citenamefont {Katsura}, \citenamefont {Shiomi},
  \citenamefont {Nagaosa},\ and\ \citenamefont {Tokura}}]{TM3}%
  \BibitemOpen
  \bibfield  {author} {\bibinfo {author} {\bibfnamefont {Y.}~\bibnamefont
  {Onose}}, \bibinfo {author} {\bibfnamefont {T.}~\bibnamefont {Ideue}},
  \bibinfo {author} {\bibfnamefont {H.}~\bibnamefont {Katsura}}, \bibinfo
  {author} {\bibfnamefont {Y.}~\bibnamefont {Shiomi}}, \bibinfo {author}
  {\bibfnamefont {N.}~\bibnamefont {Nagaosa}}, \ and\ \bibinfo {author}
  {\bibfnamefont {Y.}~\bibnamefont {Tokura}},\ }\href {\doibase
  10.1126/science.1188260} {\bibfield  {journal} {\bibinfo  {journal}
  {Science}\ }\textbf {\bibinfo {volume} {329}},\ \bibinfo {pages} {297}
  (\bibinfo {year} {2010})}\BibitemShut {NoStop}%
\bibitem [{\citenamefont {Mook}\ \emph {et~al.}(2014)\citenamefont {Mook},
  \citenamefont {Henk},\ and\ \citenamefont {Mertig}}]{TM7}%
  \BibitemOpen
  \bibfield  {author} {\bibinfo {author} {\bibfnamefont {A.}~\bibnamefont
  {Mook}}, \bibinfo {author} {\bibfnamefont {J.}~\bibnamefont {Henk}}, \ and\
  \bibinfo {author} {\bibfnamefont {I.}~\bibnamefont {Mertig}},\ }\href
  {\doibase 10.1103/PhysRevB.90.024412} {\bibfield  {journal} {\bibinfo
  {journal} {Phys. Rev. B}\ }\textbf {\bibinfo {volume} {90}},\ \bibinfo
  {pages} {024412} (\bibinfo {year} {2014})}\BibitemShut {NoStop}%
\bibitem [{\citenamefont {Kim}\ \emph {et~al.}(2016)\citenamefont {Kim},
  \citenamefont {Ochoa}, \citenamefont {Zarzuela},\ and\ \citenamefont
  {Tserkovnyak}}]{TM13}%
  \BibitemOpen
  \bibfield  {author} {\bibinfo {author} {\bibfnamefont {S.~K.}\ \bibnamefont
  {Kim}}, \bibinfo {author} {\bibfnamefont {H.}~\bibnamefont {Ochoa}}, \bibinfo
  {author} {\bibfnamefont {R.}~\bibnamefont {Zarzuela}}, \ and\ \bibinfo
  {author} {\bibfnamefont {Y.}~\bibnamefont {Tserkovnyak}},\ }\href {\doibase
  10.1103/PhysRevLett.117.227201} {\bibfield  {journal} {\bibinfo  {journal}
  {Phys. Rev. Lett.}\ }\textbf {\bibinfo {volume} {117}},\ \bibinfo {pages}
  {227201} (\bibinfo {year} {2016})}\BibitemShut {NoStop}%
\bibitem [{\citenamefont {Hidalgo-Sacoto}\ \emph {et~al.}(2020)\citenamefont
  {Hidalgo-Sacoto}, \citenamefont {Gonzalez}, \citenamefont {Vogel},
  \citenamefont {Allende}, \citenamefont {Mella}, \citenamefont {Cardenas},
  \citenamefont {Troncoso},\ and\ \citenamefont {Munoz}}]{HidalgoPRB2020}%
  \BibitemOpen
  \bibfield  {author} {\bibinfo {author} {\bibfnamefont {R.}~\bibnamefont
  {Hidalgo-Sacoto}}, \bibinfo {author} {\bibfnamefont {R.~I.}\ \bibnamefont
  {Gonzalez}}, \bibinfo {author} {\bibfnamefont {E.~E.}\ \bibnamefont {Vogel}},
  \bibinfo {author} {\bibfnamefont {S.}~\bibnamefont {Allende}}, \bibinfo
  {author} {\bibfnamefont {J.~D.}\ \bibnamefont {Mella}}, \bibinfo {author}
  {\bibfnamefont {C.}~\bibnamefont {Cardenas}}, \bibinfo {author}
  {\bibfnamefont {R.~E.}\ \bibnamefont {Troncoso}}, \ and\ \bibinfo {author}
  {\bibfnamefont {F.}~\bibnamefont {Munoz}},\ }\href {\doibase
  10.1103/PhysRevB.101.205425} {\bibfield  {journal} {\bibinfo  {journal}
  {Phys. Rev. B}\ }\textbf {\bibinfo {volume} {101}},\ \bibinfo {pages}
  {205425} (\bibinfo {year} {2020})}\BibitemShut {NoStop}%
\bibitem [{\citenamefont {Hasan}\ and\ \citenamefont {Kane}(2010)}]{TI1}%
  \BibitemOpen
  \bibfield  {author} {\bibinfo {author} {\bibfnamefont {M.~Z.}\ \bibnamefont
  {Hasan}}\ and\ \bibinfo {author} {\bibfnamefont {C.~L.}\ \bibnamefont
  {Kane}},\ }\href {\doibase 10.1103/RevModPhys.82.3045} {\bibfield  {journal}
  {\bibinfo  {journal} {Rev. Mod. Phys.}\ }\textbf {\bibinfo {volume} {82}},\
  \bibinfo {pages} {3045} (\bibinfo {year} {2010})}\BibitemShut {NoStop}%
\bibitem [{\citenamefont {Li}\ \emph {et~al.}(2009)\citenamefont {Li},
  \citenamefont {Chu}, \citenamefont {Jain},\ and\ \citenamefont
  {Shen}}]{TAI1}%
  \BibitemOpen
  \bibfield  {author} {\bibinfo {author} {\bibfnamefont {J.}~\bibnamefont
  {Li}}, \bibinfo {author} {\bibfnamefont {R.-L.}\ \bibnamefont {Chu}},
  \bibinfo {author} {\bibfnamefont {J.~K.}\ \bibnamefont {Jain}}, \ and\
  \bibinfo {author} {\bibfnamefont {S.-Q.}\ \bibnamefont {Shen}},\ }\href
  {\doibase 10.1103/PhysRevLett.102.136806} {\bibfield  {journal} {\bibinfo
  {journal} {Phys. Rev. Lett.}\ }\textbf {\bibinfo {volume} {102}},\ \bibinfo
  {pages} {136806} (\bibinfo {year} {2009})}\BibitemShut {NoStop}%
\bibitem [{\citenamefont {Groth}\ \emph {et~al.}(2009)\citenamefont {Groth},
  \citenamefont {Wimmer}, \citenamefont {Akhmerov}, \citenamefont
  {Tworzyd\l{}o},\ and\ \citenamefont {Beenakker}}]{TAI2}%
  \BibitemOpen
  \bibfield  {author} {\bibinfo {author} {\bibfnamefont {C.~W.}\ \bibnamefont
  {Groth}}, \bibinfo {author} {\bibfnamefont {M.}~\bibnamefont {Wimmer}},
  \bibinfo {author} {\bibfnamefont {A.~R.}\ \bibnamefont {Akhmerov}}, \bibinfo
  {author} {\bibfnamefont {J.}~\bibnamefont {Tworzyd\l{}o}}, \ and\ \bibinfo
  {author} {\bibfnamefont {C.~W.~J.}\ \bibnamefont {Beenakker}},\ }\href
  {\doibase 10.1103/PhysRevLett.103.196805} {\bibfield  {journal} {\bibinfo
  {journal} {Phys. Rev. Lett.}\ }\textbf {\bibinfo {volume} {103}},\ \bibinfo
  {pages} {196805} (\bibinfo {year} {2009})}\BibitemShut {NoStop}%
\bibitem [{\citenamefont {Agarwala}\ and\ \citenamefont
  {Shenoy}(2017)}]{AgarwalaPRL2017}%
  \BibitemOpen
  \bibfield  {author} {\bibinfo {author} {\bibfnamefont {A.}~\bibnamefont
  {Agarwala}}\ and\ \bibinfo {author} {\bibfnamefont {V.~B.}\ \bibnamefont
  {Shenoy}},\ }\href {\doibase 10.1103/PhysRevLett.118.236402} {\bibfield
  {journal} {\bibinfo  {journal} {Phys. Rev. Lett.}\ }\textbf {\bibinfo
  {volume} {118}},\ \bibinfo {pages} {236402} (\bibinfo {year}
  {2017})}\BibitemShut {NoStop}%
\bibitem [{\citenamefont {Costa}\ \emph {et~al.}(2019)\citenamefont {Costa},
  \citenamefont {Schleder}, \citenamefont {Buongiorno~Nardelli}, \citenamefont
  {Lewenkopf},\ and\ \citenamefont {Fazzio}}]{Costa2019}%
  \BibitemOpen
  \bibfield  {author} {\bibinfo {author} {\bibfnamefont {M.}~\bibnamefont
  {Costa}}, \bibinfo {author} {\bibfnamefont {G.~R.}\ \bibnamefont {Schleder}},
  \bibinfo {author} {\bibfnamefont {M.}~\bibnamefont {Buongiorno~Nardelli}},
  \bibinfo {author} {\bibfnamefont {C.}~\bibnamefont {Lewenkopf}}, \ and\
  \bibinfo {author} {\bibfnamefont {A.}~\bibnamefont {Fazzio}},\ }\href
  {\doibase 10.1021/acs.nanolett.9b03881} {\bibfield  {journal} {\bibinfo
  {journal} {Nano Letters}\ }\textbf {\bibinfo {volume} {19}},\ \bibinfo
  {pages} {8941–8946} (\bibinfo {year} {2019})}\BibitemShut {NoStop}%
\bibitem [{\citenamefont {Marsal}\ \emph {et~al.}(2020)\citenamefont {Marsal},
  \citenamefont {Varjas},\ and\ \citenamefont {Grushin}}]{Marsal2020}%
  \BibitemOpen
  \bibfield  {author} {\bibinfo {author} {\bibfnamefont {Q.}~\bibnamefont
  {Marsal}}, \bibinfo {author} {\bibfnamefont {D.}~\bibnamefont {Varjas}}, \
  and\ \bibinfo {author} {\bibfnamefont {A.~G.}\ \bibnamefont {Grushin}},\
  }\href {\doibase 10.1073/pnas.2007384117} {\bibfield  {journal} {\bibinfo
  {journal} {Proceedings of the National Academy of Sciences}\ }\textbf
  {\bibinfo {volume} {117}},\ \bibinfo {pages} {30260–30265} (\bibinfo {year}
  {2020})}\BibitemShut {NoStop}%
\bibitem [{\citenamefont {Bianco}\ and\ \citenamefont
  {Resta}(2011)}]{BiancoPRB2011}%
  \BibitemOpen
  \bibfield  {author} {\bibinfo {author} {\bibfnamefont {R.}~\bibnamefont
  {Bianco}}\ and\ \bibinfo {author} {\bibfnamefont {R.}~\bibnamefont {Resta}},\
  }\href {\doibase 10.1103/PhysRevB.84.241106} {\bibfield  {journal} {\bibinfo
  {journal} {Phys. Rev. B}\ }\textbf {\bibinfo {volume} {84}},\ \bibinfo
  {pages} {241106} (\bibinfo {year} {2011})}\BibitemShut {NoStop}%
\bibitem [{\citenamefont {Cerjan}\ and\ \citenamefont
  {Loring}(2022)}]{CerjanPRB2022}%
  \BibitemOpen
  \bibfield  {author} {\bibinfo {author} {\bibfnamefont {A.}~\bibnamefont
  {Cerjan}}\ and\ \bibinfo {author} {\bibfnamefont {T.~A.}\ \bibnamefont
  {Loring}},\ }\href {\doibase 10.1103/PhysRevB.106.064109} {\bibfield
  {journal} {\bibinfo  {journal} {Phys. Rev. B}\ }\textbf {\bibinfo {volume}
  {106}},\ \bibinfo {pages} {064109} (\bibinfo {year} {2022})}\BibitemShut
  {NoStop}%
\bibitem [{\citenamefont {Loring}\ and\ \citenamefont
  {Hastings}(2010)}]{Bott3}%
  \BibitemOpen
  \bibfield  {author} {\bibinfo {author} {\bibfnamefont {T.~A.}\ \bibnamefont
  {Loring}}\ and\ \bibinfo {author} {\bibfnamefont {M.~B.}\ \bibnamefont
  {Hastings}},\ }\href {\doibase 10.1209/0295-5075/92/67004} {\bibfield
  {journal} {\bibinfo  {journal} {{EPL} (Europhysics Letters)}\ }\textbf
  {\bibinfo {volume} {92}},\ \bibinfo {pages} {67004} (\bibinfo {year}
  {2010})}\BibitemShut {NoStop}%
\bibitem [{\citenamefont {Toniolo}(2022)}]{Toniolo2022}%
  \BibitemOpen
  \bibfield  {author} {\bibinfo {author} {\bibfnamefont {D.}~\bibnamefont
  {Toniolo}},\ }\href {http://dx.doi.org/10.1007/s11005-022-01602-6} {\bibfield
   {journal} {\bibinfo  {journal} {Letters in Mathematical Physics}\ }\textbf
  {\bibinfo {volume} {112}} (\bibinfo {year} {2022})}\BibitemShut {NoStop}%
\bibitem [{\citenamefont {Wang}\ \emph {et~al.}(2020)\citenamefont {Wang},
  \citenamefont {Brataas},\ and\ \citenamefont {Troncoso}}]{WangPRL2020}%
  \BibitemOpen
  \bibfield  {author} {\bibinfo {author} {\bibfnamefont {X.~S.}\ \bibnamefont
  {Wang}}, \bibinfo {author} {\bibfnamefont {A.}~\bibnamefont {Brataas}}, \
  and\ \bibinfo {author} {\bibfnamefont {R.~E.}\ \bibnamefont {Troncoso}},\
  }\href {\doibase 10.1103/PhysRevLett.125.217202} {\bibfield  {journal}
  {\bibinfo  {journal} {Phys. Rev. Lett.}\ }\textbf {\bibinfo {volume} {125}},\
  \bibinfo {pages} {217202} (\bibinfo {year} {2020})}\BibitemShut {NoStop}%
\bibitem [{\citenamefont {Liu}\ \emph {et~al.}(2006)\citenamefont {Liu},
  \citenamefont {Sellmyer},\ and\ \citenamefont {Shindo}}]{liu2006handbook}%
  \BibitemOpen
  \bibfield  {author} {\bibinfo {author} {\bibfnamefont {Y.}~\bibnamefont
  {Liu}}, \bibinfo {author} {\bibfnamefont {D.~J.}\ \bibnamefont {Sellmyer}}, \
  and\ \bibinfo {author} {\bibfnamefont {D.}~\bibnamefont {Shindo}},\
  }\href@noop {} {\emph {\bibinfo {title} {Handbook of Advanced Magnetic
  Materials: Vol 1. Nanostructural Effects. Vol 2. Characterization and
  Simulation. Vol 3. Fabrication and Processing. Vol 4. Properties and
  Applications}}}\ (\bibinfo  {publisher} {Springer},\ \bibinfo {year}
  {2006})\BibitemShut {NoStop}%
\bibitem [{Note1()}]{Note1}%
  \BibitemOpen
  \bibinfo {note} {In presence of disorder the shape of each skyrmion is
  distorted and thus, we assume for their positions the location of the spin
  pointing downwards in the magnetic structure.}\BibitemShut {Stop}%
\bibitem [{\citenamefont {Nishikawa}\ \emph {et~al.}(2019)\citenamefont
  {Nishikawa}, \citenamefont {Hukushima},\ and\ \citenamefont
  {Krauth}}]{nishikawa2019solid}%
  \BibitemOpen
  \bibfield  {author} {\bibinfo {author} {\bibfnamefont {Y.}~\bibnamefont
  {Nishikawa}}, \bibinfo {author} {\bibfnamefont {K.}~\bibnamefont
  {Hukushima}}, \ and\ \bibinfo {author} {\bibfnamefont {W.}~\bibnamefont
  {Krauth}},\ }\href {https://doi.org/10.1103/PhysRevB.99.064435} {\bibfield
  {journal} {\bibinfo  {journal} {Phys. Rev. B}\ }\textbf {\bibinfo {volume}
  {99}},\ \bibinfo {pages} {064435} (\bibinfo {year} {2019})}\BibitemShut
  {NoStop}%
\bibitem [{\citenamefont {Z{\'a}zvorka}\ \emph {et~al.}(2020)\citenamefont
  {Z{\'a}zvorka}, \citenamefont {Dittrich}, \citenamefont {Ge}, \citenamefont
  {Kerber}, \citenamefont {Raab}, \citenamefont {Winkler}, \citenamefont
  {Litzius}, \citenamefont {Veis}, \citenamefont {Virnau},\ and\ \citenamefont
  {Kl{\"a}ui}}]{zazvorka2020skyrmion}%
  \BibitemOpen
  \bibfield  {author} {\bibinfo {author} {\bibfnamefont {J.}~\bibnamefont
  {Z{\'a}zvorka}}, \bibinfo {author} {\bibfnamefont {F.}~\bibnamefont
  {Dittrich}}, \bibinfo {author} {\bibfnamefont {Y.}~\bibnamefont {Ge}},
  \bibinfo {author} {\bibfnamefont {N.}~\bibnamefont {Kerber}}, \bibinfo
  {author} {\bibfnamefont {K.}~\bibnamefont {Raab}}, \bibinfo {author}
  {\bibfnamefont {T.}~\bibnamefont {Winkler}}, \bibinfo {author} {\bibfnamefont
  {K.}~\bibnamefont {Litzius}}, \bibinfo {author} {\bibfnamefont
  {M.}~\bibnamefont {Veis}}, \bibinfo {author} {\bibfnamefont {P.}~\bibnamefont
  {Virnau}}, \ and\ \bibinfo {author} {\bibfnamefont {M.}~\bibnamefont
  {Kl{\"a}ui}},\ }\href {https://doi.org/10.1002/adfm.202004037} {\bibfield
  {journal} {\bibinfo  {journal} {Advanced Functional Materials}\ }\textbf
  {\bibinfo {volume} {30}},\ \bibinfo {pages} {2004037} (\bibinfo {year}
  {2020})}\BibitemShut {NoStop}%
\bibitem [{\citenamefont {Raftrey}\ \emph {et~al.}(2023)\citenamefont
  {Raftrey}, \citenamefont {Chen}, \citenamefont {Shao}, \citenamefont {Chan},
  \citenamefont {Yalisove}, \citenamefont {Chen}, \citenamefont {Yao},
  \citenamefont {Scott}, \citenamefont {Wu}, \citenamefont {Muller} \emph
  {et~al.}}]{raftrey2023ordering}%
  \BibitemOpen
  \bibfield  {author} {\bibinfo {author} {\bibfnamefont {D.}~\bibnamefont
  {Raftrey}}, \bibinfo {author} {\bibfnamefont {X.}~\bibnamefont {Chen}},
  \bibinfo {author} {\bibfnamefont {Y.-T.}\ \bibnamefont {Shao}}, \bibinfo
  {author} {\bibfnamefont {Y.-T.}\ \bibnamefont {Chan}}, \bibinfo {author}
  {\bibfnamefont {R.}~\bibnamefont {Yalisove}}, \bibinfo {author}
  {\bibfnamefont {R.}~\bibnamefont {Chen}}, \bibinfo {author} {\bibfnamefont
  {J.}~\bibnamefont {Yao}}, \bibinfo {author} {\bibfnamefont {M.}~\bibnamefont
  {Scott}}, \bibinfo {author} {\bibfnamefont {W.}~\bibnamefont {Wu}}, \bibinfo
  {author} {\bibfnamefont {D.}~\bibnamefont {Muller}},  \emph {et~al.},\
  }\href@noop {} {\bibfield  {journal} {\bibinfo  {journal} {Nature
  Communications}\ }\textbf {\bibinfo {volume} {14}} (\bibinfo {year}
  {2023})}\BibitemShut {NoStop}%
\bibitem [{\citenamefont {Holstein}\ and\ \citenamefont
  {Primakoff}(1940)}]{HolsteinPR1940}%
  \BibitemOpen
  \bibfield  {author} {\bibinfo {author} {\bibfnamefont {T.}~\bibnamefont
  {Holstein}}\ and\ \bibinfo {author} {\bibfnamefont {H.}~\bibnamefont
  {Primakoff}},\ }\href {\doibase 10.1103/PhysRev.58.1098} {\bibfield
  {journal} {\bibinfo  {journal} {Phys. Rev.}\ }\textbf {\bibinfo {volume}
  {58}},\ \bibinfo {pages} {1098} (\bibinfo {year} {1940})}\BibitemShut
  {NoStop}%
\bibitem [{\citenamefont {Colpa}(1978)}]{colpa1978}%
  \BibitemOpen
  \bibfield  {author} {\bibinfo {author} {\bibfnamefont {J.}~\bibnamefont
  {Colpa}},\ }\href@noop {} {\bibfield  {journal} {\bibinfo  {journal} {Physica
  A: Statistical Mechanics and its Applications}\ }\textbf {\bibinfo {volume}
  {93}},\ \bibinfo {pages} {327} (\bibinfo {year} {1978})}\BibitemShut
  {NoStop}%
\bibitem [{Note2()}]{Note2}%
  \BibitemOpen
  \bibinfo {note} {For fermionic systems the metric $\zeta $ reduces to the
  identity and the definition in the main text returns to the electronic Bott
  index}\BibitemShut {NoStop}%
\bibitem [{\citenamefont {Rostami}\ and\ \citenamefont
  {Cappelluti}(2017)}]{RostamiPRB2017}%
  \BibitemOpen
  \bibfield  {author} {\bibinfo {author} {\bibfnamefont {H.}~\bibnamefont
  {Rostami}}\ and\ \bibinfo {author} {\bibfnamefont {E.}~\bibnamefont
  {Cappelluti}},\ }\href {https://link.aps.org/doi/10.1103/PhysRevB.96.054205}
  {\bibfield  {journal} {\bibinfo  {journal} {Phys. Rev. B}\ }\textbf {\bibinfo
  {volume} {96}},\ \bibinfo {pages} {054205} (\bibinfo {year}
  {2017})}\BibitemShut {NoStop}%
\bibitem [{\citenamefont {Chisnell}\ \emph {et~al.}(2015)\citenamefont
  {Chisnell}, \citenamefont {Helton}, \citenamefont {Freedman}, \citenamefont
  {Singh}, \citenamefont {Bewley}, \citenamefont {Nocera},\ and\ \citenamefont
  {Lee}}]{chisnell2015topological}%
  \BibitemOpen
  \bibfield  {author} {\bibinfo {author} {\bibfnamefont {R.}~\bibnamefont
  {Chisnell}}, \bibinfo {author} {\bibfnamefont {J.~S.}\ \bibnamefont
  {Helton}}, \bibinfo {author} {\bibfnamefont {D.~E.}\ \bibnamefont
  {Freedman}}, \bibinfo {author} {\bibfnamefont {D.~K.}\ \bibnamefont {Singh}},
  \bibinfo {author} {\bibfnamefont {R.~I.}\ \bibnamefont {Bewley}}, \bibinfo
  {author} {\bibfnamefont {D.~G.}\ \bibnamefont {Nocera}}, \ and\ \bibinfo
  {author} {\bibfnamefont {Y.~S.}\ \bibnamefont {Lee}},\ }\href {\doibase
  10.1103/PhysRevLett.115.147201} {\bibfield  {journal} {\bibinfo  {journal}
  {Phys. Rev. Lett.}\ }\textbf {\bibinfo {volume} {115}},\ \bibinfo {pages}
  {147201} (\bibinfo {year} {2015})}\BibitemShut {NoStop}%
\bibitem [{\citenamefont {Vi\~nas Bostr\"om}\ \emph {et~al.}(2023)\citenamefont
  {Vi\~nas Bostr\"om}, \citenamefont {Parvini}, \citenamefont {McIver},
  \citenamefont {Rubio}, \citenamefont {Kusminskiy},\ and\ \citenamefont
  {Sentef}}]{VinasPRL2022}%
  \BibitemOpen
  \bibfield  {author} {\bibinfo {author} {\bibfnamefont {E.}~\bibnamefont
  {Vi\~nas Bostr\"om}}, \bibinfo {author} {\bibfnamefont {T.~S.}\ \bibnamefont
  {Parvini}}, \bibinfo {author} {\bibfnamefont {J.~W.}\ \bibnamefont {McIver}},
  \bibinfo {author} {\bibfnamefont {A.}~\bibnamefont {Rubio}}, \bibinfo
  {author} {\bibfnamefont {S.~V.}\ \bibnamefont {Kusminskiy}}, \ and\ \bibinfo
  {author} {\bibfnamefont {M.~A.}\ \bibnamefont {Sentef}},\ }\href {\doibase
  10.1103/PhysRevLett.130.026701} {\bibfield  {journal} {\bibinfo  {journal}
  {Phys. Rev. Lett.}\ }\textbf {\bibinfo {volume} {130}},\ \bibinfo {pages}
  {026701} (\bibinfo {year} {2023})}\BibitemShut {NoStop}%
\bibitem [{\citenamefont {Zhou}\ \emph {et~al.}(2021)\citenamefont {Zhou},
  \citenamefont {Carmiggelt}, \citenamefont {G\"{a}chter}, \citenamefont
  {Esterlis}, \citenamefont {Sels}, \citenamefont {St\"{o}hr}, \citenamefont
  {Du}, \citenamefont {Fernandez}, \citenamefont {Rodriguez-Nieva},
  \citenamefont {B\"{u}ttner}, \citenamefont {Demler},\ and\ \citenamefont
  {Yacoby}}]{Zhou2021}%
  \BibitemOpen
  \bibfield  {author} {\bibinfo {author} {\bibfnamefont {T.~X.}\ \bibnamefont
  {Zhou}}, \bibinfo {author} {\bibfnamefont {J.~J.}\ \bibnamefont
  {Carmiggelt}}, \bibinfo {author} {\bibfnamefont {L.~M.}\ \bibnamefont
  {G\"{a}chter}}, \bibinfo {author} {\bibfnamefont {I.}~\bibnamefont
  {Esterlis}}, \bibinfo {author} {\bibfnamefont {D.}~\bibnamefont {Sels}},
  \bibinfo {author} {\bibfnamefont {R.~J.}\ \bibnamefont {St\"{o}hr}}, \bibinfo
  {author} {\bibfnamefont {C.}~\bibnamefont {Du}}, \bibinfo {author}
  {\bibfnamefont {D.}~\bibnamefont {Fernandez}}, \bibinfo {author}
  {\bibfnamefont {J.~F.}\ \bibnamefont {Rodriguez-Nieva}}, \bibinfo {author}
  {\bibfnamefont {F.}~\bibnamefont {B\"{u}ttner}}, \bibinfo {author}
  {\bibfnamefont {E.}~\bibnamefont {Demler}}, \ and\ \bibinfo {author}
  {\bibfnamefont {A.}~\bibnamefont {Yacoby}},\ }\href
  {http://dx.doi.org/10.1073/pnas.2019473118} {\bibfield  {journal} {\bibinfo
  {journal} {Proceedings of the National Academy of Sciences}\ }\textbf
  {\bibinfo {volume} {118}} (\bibinfo {year} {2021})}\BibitemShut {NoStop}%
\bibitem [{\citenamefont {Hrabec}\ \emph {et~al.}(2017)\citenamefont {Hrabec},
  \citenamefont {Sampaio}, \citenamefont {Belmeguenai}, \citenamefont {Gross},
  \citenamefont {Weil}, \citenamefont {Ch{\'e}rif}, \citenamefont
  {Stashkevich}, \citenamefont {Jacques}, \citenamefont {Thiaville},\ and\
  \citenamefont {Rohart}}]{hrabec2017current}%
  \BibitemOpen
  \bibfield  {author} {\bibinfo {author} {\bibfnamefont {A.}~\bibnamefont
  {Hrabec}}, \bibinfo {author} {\bibfnamefont {J.}~\bibnamefont {Sampaio}},
  \bibinfo {author} {\bibfnamefont {M.}~\bibnamefont {Belmeguenai}}, \bibinfo
  {author} {\bibfnamefont {I.}~\bibnamefont {Gross}}, \bibinfo {author}
  {\bibfnamefont {R.}~\bibnamefont {Weil}}, \bibinfo {author} {\bibfnamefont
  {S.~M.}\ \bibnamefont {Ch{\'e}rif}}, \bibinfo {author} {\bibfnamefont
  {A.}~\bibnamefont {Stashkevich}}, \bibinfo {author} {\bibfnamefont
  {V.}~\bibnamefont {Jacques}}, \bibinfo {author} {\bibfnamefont
  {A.}~\bibnamefont {Thiaville}}, \ and\ \bibinfo {author} {\bibfnamefont
  {S.}~\bibnamefont {Rohart}},\ }\href {https://doi.org/10.1038/ncomms15765}
  {\bibfield  {journal} {\bibinfo  {journal} {Nature communications}\ }\textbf
  {\bibinfo {volume} {8}},\ \bibinfo {pages} {15765} (\bibinfo {year}
  {2017})}\BibitemShut {NoStop}%
\bibitem [{\citenamefont {Juge}\ \emph {et~al.}(2019)\citenamefont {Juge},
  \citenamefont {Je}, \citenamefont {de~Souza~Chaves}, \citenamefont
  {Buda-Prejbeanu}, \citenamefont {Pe{\~n}a-Garcia}, \citenamefont {Nath},
  \citenamefont {Miron}, \citenamefont {Rana}, \citenamefont {Aballe},
  \citenamefont {Foerster} \emph {et~al.}}]{juge2019current}%
  \BibitemOpen
  \bibfield  {author} {\bibinfo {author} {\bibfnamefont {R.}~\bibnamefont
  {Juge}}, \bibinfo {author} {\bibfnamefont {S.-G.}\ \bibnamefont {Je}},
  \bibinfo {author} {\bibfnamefont {D.}~\bibnamefont {de~Souza~Chaves}},
  \bibinfo {author} {\bibfnamefont {L.~D.}\ \bibnamefont {Buda-Prejbeanu}},
  \bibinfo {author} {\bibfnamefont {J.}~\bibnamefont {Pe{\~n}a-Garcia}},
  \bibinfo {author} {\bibfnamefont {J.}~\bibnamefont {Nath}}, \bibinfo {author}
  {\bibfnamefont {I.~M.}\ \bibnamefont {Miron}}, \bibinfo {author}
  {\bibfnamefont {K.~G.}\ \bibnamefont {Rana}}, \bibinfo {author}
  {\bibfnamefont {L.}~\bibnamefont {Aballe}}, \bibinfo {author} {\bibfnamefont
  {M.}~\bibnamefont {Foerster}},  \emph {et~al.},\ }\href
  {https://doi.org/10.1103/PhysRevApplied.12.044007} {\bibfield  {journal}
  {\bibinfo  {journal} {Physical Review Applied}\ }\textbf {\bibinfo {volume}
  {12}},\ \bibinfo {pages} {044007} (\bibinfo {year} {2019})}\BibitemShut
  {NoStop}%
\end{thebibliography}%

\end{document}